\documentclass{article}
\usepackage{cite}
\usepackage{amsmath,amssymb,amsfonts}
\usepackage{algorithmic}
\usepackage{graphicx}
\usepackage{textcomp}
\usepackage{wrapfig}
\usepackage[ruled]{algorithm2e}
\usepackage{caption}
\usepackage{subcaption}
\usepackage{multirow}

\usepackage[a4paper, left=2cm, right=2cm, top=2.5cm, bottom=2.5cm]{geometry}

\usepackage{soul} 
\usepackage{xcolor}
\usepackage{dblfloatfix}
\usepackage{booktabs}
\usepackage{wrapfig}
\usepackage{cite}
\usepackage{float}
\usepackage{listings}
\usepackage{calc}
\usepackage{enumitem}
\usepackage{diagbox}
\usepackage{textgreek}
\usepackage{changepage,threeparttable} 
\usepackage{makecell}
\usepackage{amssymb}
\usepackage{hyperref}

\definecolor{lightblue}{rgb}{0.68, 0.85, 0.9} 
\definecolor{seagreen}{rgb}{0.18, 0.42, 0.41}
\definecolor{lightgreen}{rgb}{0.596, 0.733, 0.525} 
\definecolor{lightyellow}{rgb}{0.961, 0.914, 0.765} 
\definecolor{lighterblue}{rgb}{0.824, 0.894, 0.953} 
\definecolor{darkbrown}{rgb}{0.550, 0.094, 0.067} 

\newcommand{\highlightyellow}[1]{\sethlcolor{lightyellow}\hl{#1}}
\newcommand{\highlightgreen}[1]{\sethlcolor{lightgreen}\hl{#1}}
\newcommand{\highlightblue}[1]{\sethlcolor{lighterblue}\hl{#1}}
\newcommand{\highlightbrown}[1]{\sethlcolor{darkbrown}\hl{#1}}

\usepackage{microtype}
\usepackage[utf8]{inputenc}
\usepackage{comment}
\usepackage{diagbox}
\usepackage{cancel}
\usepackage{lipsum} 
\newcommand{\Eq}{Eq.}
\usepackage{adjustbox} 
\usepackage{tablefootnote}

\usepackage{authblk}

{\lccode`\?=`\p \lccode`\!=`\t  \lowercase{\gdef\ignorept#1?!{#1}}}
\def\divbyccvv#1{\expandafter\ignorept\the\dimexpr#1pt/255\relax}

\def\defineCMYKcolor#1#2{\defineCMYKcolorA{#1}#2,} 
\def\defineCMYKcolorA#1#2,#3,#4,#5,{\edef\tmp{\noexpand\definecolor{#1}{cmyk}%
		{\divbyccvv{#2},\divbyccvv{#3},\divbyccvv{#4},\divbyccvv{#5}}}\tmp
}

\defineCMYKcolor{unimint}{0,0,0,0}

\newcommand{\mandel}{\mbox{Mandelbrot\xspace}}
\newcommand{\streamtriad}{\mbox{STREAM Triad\xspace}}
\newcommand{\openmp}{\mbox{Open{MP}\xspace}}
\newcommand{\lbomp}{\mbox{LB4OMP\xspace}}

\newcommand{\sphynx}{\mbox{SPHYNX Evrard collapse\xspace}}
\newcommand{\tc}{\mbox{TC\xspace}}
\newcommand{\lulesh}{\mbox{LULESH\xspace}}
\newcommand{\hacc}{\mbox{HACCKernels\xspace}}

\newcommand{\Auto}{\mbox{Auto4OMP\xspace}}
\newcommand{\rlomp}{\mbox{RL4OMP\xspace}}

\newcommand{\static}{\mbox{\texttt{STATIC~}\xspace}}
\newcommand{\ssdynamic}{\mbox{\texttt{SS~}\xspace}}
\newcommand{\gss}{\mbox{\texttt{GSS~}\xspace}}
\newcommand{\llvmauto}{\mbox{\texttt{Auto(LLVM)}\xspace}}
\newcommand{\tss}{\mbox{\texttt{TSS}\xspace}}
\newcommand{\steal}{\mbox{\texttt{Static Steal}\xspace}}
\newcommand{\factwo}{\mbox{\texttt{mFAC2}\xspace}}
\newcommand{\awfb}{\texttt{\mbox{AWF-B}\xspace}}
\newcommand{\awfc}{\texttt{\mbox{AWF-C}\xspace}}
\newcommand{\awfd}{\texttt{\mbox{AWF-D}\xspace}}
\newcommand{\awfe}{\texttt{\mbox{AWF-E}\xspace}}
\newcommand{\af}{\mbox{\texttt{mAF}\xspace}}

\newcommand{\gt}{\mbox{\texttt{Oracle}\xspace}}

\newcommand{\autoRND}{\mbox{\texttt{RandomSel}\xspace}}
\newcommand{\autoEXT}{\mbox{\texttt{ExhaustiveSel}\xspace}}

\newcommand{\autoEXP}{\mbox{\texttt{ExpertSel}\xspace}}

\newcommand{\qlearn}{\mbox{\texttt{Q-Learn}\xspace}}
\newcommand{\sarsa}{\mbox{\texttt{SARSA}\xspace}}

\newcommand{\expchunk}{\mbox{\texttt{expChunk}\xspace}}

\newcommand{\xeon}{\mbox{Broadwell~\xspace}}
\newcommand{\gpu}{\mbox{Cascade-Lake~\xspace}}
\newcommand{\amd}{\mbox{EPYC~\xspace}}

\newcommand{\cov}{\emph{c.o.v.}\xspace}
\definecolor{darkgreen}{RGB}{17, 89, 9}
\definecolor{magentaFig3}{RGB}{255, 23, 227}
\definecolor{blueFig3}{RGB}{133, 206, 219}

\usepackage{bm}

\sethlcolor{white}

\def\BibTeX{{\rm B\kern-.05em{\sc i\kern-.025em b}\kern-.08em
		T\kern-.1667em\lower.7ex\hbox{E}\kern-.125emX}}

\title{A Comparative Study of OpenMP Scheduling Algorithm Selection Strategies}

	\author[1]{Jonas H. M\"uller Kornd\"orfer}
	\author[2]{Ali Mohammed}
	\author[2]{Ahmed Eleliemy}
	\author[1]{Quentin Guilloteau}
        \author[1]{Reto Krummenacher}
        \author[1]{Florina M. Ciorba}
	\affil[1]{
		Department of Mathematics and Computer Science\\

		University of Basel, Basel, Switzerland\\

		Email: jonas.korndorfer@unibas.ch, quentin.guilloteau@unibas.ch, reto.krummenacher@unibas.ch, and florina.ciorba@unibas.ch
	}
	\affil[2]{
		HPE HPC/AI EMEA Research Lab\\
            
            Basel, Switzerland \\

		Email: ali.mohammed@hpe.com, ahmed.eleliemy@hpe.com
	}
\begin{document}

\sloppy

\maketitle

	\begin{abstract}
		Scientific and data science applications are becoming more complex, with increasingly demanding computational and memory requirements during execution. 
		Modern high performance computing (HPC) systems offer increased parallelism and heterogeneity across nodes, devices, and cores. 
		Thus, effective scheduling and load balancing techniques are crucial to maximize applications performance on such systems.
		Commonly used node-level parallelization frameworks, such as OpenMP, employ an increasing number of advanced scheduling algorithms to support various applications and HPC platforms. 
		For arbitrary application-system pairs, this results in an instance of the \textit{scheduling algorithm selection problem}, which requires fast and accurate selection methods for
		the diversity of applications' workloads and computing systems' characteristics. 
		In this work, we study the problem of learning to select scheduling algorithms in OpenMP. 
		Specifically, we propose and use expert-based and reinforcement learning (RL) based selection approaches.  
		We assess the effectiveness of each approach through an extensive performance analysis campaign with six applications and three systems, exposing capabilities and limits. 
		We found that learning the highest-performing scheduling algorithm using RL-based methods is effective, but at a high exploration cost, and that the most relevant factor is the type of reward used by the RL methods.
		We also found that expert-based selection indeed leverages expert knowledge with fewer exploration needs, albeit at the risk of not selecting the highest-performing scheduling algorithm for a given application-system pair. 
		We combine expert knowledge with RL-based approaches and show improved performance.
		Overall, this work shows that the selection of scheduling algorithms during execution is possible and beneficial for OpenMP applications. 
		We anticipate that this work can be extended and combined with the selection of scheduling algorithms for MPI-based applications (provided that there is a portfolio of scheduling algorithms on distributed memory nodes), thereby optimizing scheduling decisions across parallelism levels. 
	\end{abstract}

\section{Introduction}

Modern HPC systems offer significant computational power for scientific and data analysis applications. Their complexity arises from enhanced parallelism and heterogeneity at various levels. 
	Performance can vary due to factors like code branching and data access patterns, leading to uneven completion times between nodes and cores. 
	This can result in load imbalance and significant performance degradation.

	Efficient scheduling algorithms are essential to balance the load and maximize the performance of HPC applications. 
	Recent work highlights the need to integrate various scheduling algorithms into frameworks like OpenMP~\cite{ciorba:2018,penna2019comprehensive, Kasielke:2019, lb4omptpds}. 
	This creates numerous options for any application-system pair, complicating the choice of the optimal strategy, especially in time-stepping scenarios. 
	The varying computational demands and system capabilities, combined with many scheduling choices, exemplify Rice's algorithm selection problem~\cite{RICE1976}, which aims to find the highest-performing algorithm for specific application and system pairings.

	Recently, we proposed scheduling algorithm selection based on expert knowledge to solve the algorithm selection problem for OpenMP applications~\cite{auto4omp}. 
	This approach, called~\Auto, involved the implementation of three new methods—\autoRND, \autoEXT, and \autoEXP{}—as extensions to the LB4OMP library~\cite{lb4omptpds}. 
	The selection portfolio comprises 12 scheduling algorithms from LB4OMP, which is an extension to the LLVM OpenMP runtime.

	Expert-based selection methods rely on expert knowledge to define rules to choose a scheduling algorithm from a portfolio. 
	Integrating new algorithms into this portfolio involves: 
	1)~understanding the new algorithm to properly revise the selection rules, and 
	2)~modifying the selection method's source code. 
	Additionally, obtaining the necessary expert knowledge often requires costly, time-consuming, or impractical experimentation across various applications, loops, and systems, which can result in inefficient algorithm selection.

	In this work, in an attempt to address these shortcomings, we propose a reinforcement learning (RL)-based extension to LB4OMP and~\Auto{} for automated online selection of scheduling algorithms in OpenMP applications. 
	We implemented and adapted the model-free \qlearn~\cite{qlearn} and \sarsa~\cite{suttonReinforcement1998} RL algorithms for this purpose.
	A preliminary version of this work appeared in 2023~\cite{rl4ompFlorina2023}.

	We conducted a comparative study to evaluate the effectiveness and performance of expert- and RL-based selection methods by analyzing their strengths and weaknesses. 
	This involved an extensive performance analysis campaign, executing six applications with unique compute/memory-bound and workload imbalance characteristics across three computing systems in 720 configuration combinations, totaling 3'600 executions.

	We show that RL-based methods effectively identify the highest-performing scheduling algorithms, though they incur significant exploration overhead. 
	We identified that the choice of reward metric is crucial for algorithm selection with these RL methods. 
	In contrast, expert-based selection minimizes the need for extensive exploration by leveraging expert knowledge, but it risks not selecting the optimal scheduling algorithm for specific application-system pairs.
	In addition, we show that integrating expert knowledge with RL-based methods can enhance their performance.

	\textbf{Contributions.}
	This work assesses and advocates the automated selection of scheduling algorithms in OpenMP multithreaded applications to better exploit node-level parallelism. 
	We focus on improving adaptability and performance portability using expert- and RL-based selection methods. 
	Specifically, the contributions of this study include the following: 
	\begin{enumerate}[align=left, leftmargin=*]
		\item \emph{The proposal and implementation} of two RL-based methods for automated scheduling algorithm selection for OpenMP loops.
		\item \emph{An in-depth performance analysis campaign} evaluating the potential and limitations of both expert- and RL-based automated selection methods.
		\item \emph{Recommendations} on scenarios with outperforming methods and strategies for integrating expert knowledge with RL-based methods to further improve performance.
	\end{enumerate}
	
	This work is organized as follows: 
	Sect.~\ref{sec:back} provides the necessary background regarding all scheduling algorithms approached in this paper.
	Sect.~\ref{sec:main} outlines the research questions and details the design, implementation, and use of the proposed RL-based and existing expert-based~\cite{auto4omp} selection methods. 
	Sect.~\ref{sec:perfanalysis} presents a comparative study of these methods and addresses research questions. 
	Sect.~\ref{sec:rlw} reviews the related literature. 
	Sect.~\ref{sec:conclusion} discusses limitations, concludes the work, and suggests future research directions.

\section{Background}\label{sec:back}
	\hl{This section provides the necessary background on the scheduling algorithms used in this study. 
		It offers insights into how different scheduling techniques approach the distribution of loop iterations and describes both the relevance and operational aspects of the OpenMP chunk parameter. 
		Understanding these mechanisms is essential to appreciate their impact on performance and load balancing in the context of this work.}

	\hl{We define in \mbox{Table~\ref{tab:notationsched}} the symbols used to describe how different scheduling techniques calculate different chunk sizes (chunks of loop iterations).
		In loop scheduling, a loop iteration refers to the smallest unit of work possible while a \textbf{chunk} of iterations represents a certain number of iterations. 
		The chunk size calculated by a scheduling algorithm represents the number of iterations that a given thread or process will receive per work request. 
		Note that for most \textit{dynamic} scheduling algorithms the chunk size varies along the execution of a loop, usually starting with large chunks of iterations, followed by smaller chunks, and then down to single iterations (in case no limiter is specified, e.g. a chunk parameter).}

	\begin{table}[!htb]
		\centering
		\caption{\hl{Notation used to describe the loop scheduling algorithms and their chunk size calculation.}}
		\label{tab:notationsched}
		\begin{tabular}{@{}l|l@{}}
			\textbf{Symbol} & \textbf{Description}                     \\ \hline 
			$P$ & The number of processing elements (PEs).\\
			$P_{i}$ & Processing element of id $i$.\\
			$w$ & Weight of a given PE.\\
			$N$ & The number of iterations.\\
			$Cs$ & Chunk size.\\
			$Cs_{i}$ & Chunk size on a given $i$ scheduling round.\\
			$l$ & Last chunk size.\\
			$f$ & First chunk size.\\
			$b$ & Batch of chunks.\\
			$j$ & Id of a batch of chunks.\\
			$A$ & Number of chunks. \\
			$R$ & The number of remaining iterations for a given loop.\\
			$R_{i}$ & Number of remaining iterations for a given $i$th chunk.\\
			$\mu$ & Mean of iterations execution times.\\
			$\sigma$ & Standard deviation of iterations execution times.\\
			$h$ & The scheduling overhead of assigning loop iterations.\\
		\end{tabular}
	\end{table}

	\hl{
		\noindent\textbf{Significance of the chunk parameter.}\ 
		Both the standard scheduling algorithms in \openmp{} and those in \lbomp{} allow users to specify a \textit{chunk parameter}.
		The chunk parameter works differently depending on the scheduling algorithm. 
		For \mbox{\texttt{schedule(static, chunk)}} and \mbox{\texttt{schedule(dynamic,chunk)}}, the \textit{chunk parameter} directly determines the number of iterations each thread receives with every work request. 
		For all other scheduling algorithms, the \textit{chunk parameter} serves as a \textbf{threshold}: if the chunk size calculated by the algorithm drops below the \textit{chunk parameter} value, it is set to match the chunk parameter instead:
		chunk$_\text{delivered}$ = max(chunk$_\text{algorithm}$, chunk$_\text{user}$).
	}
	
	\hl{
		The \texttt{chunk} parameter was introduced into the \openmp{} standard as a means to reduce scheduling overhead and improve data locality.
		Carefully choosing a suitable \textit{chunk parameter} can boost performance, as it leads to fewer scheduling rounds per thread compared to not setting a threshold (the default chunk parameter or minimum chunk is 1 loop iteration in most OpenMP implementations, including the LLVM base used in this work).
		Detailed experimental results exploring the potential of numerous chunk parameters combined with several different scheduling algorithms can be found in related \mbox{work~\cite{lb4omptpds}}.
	}
	
	\hl{
		Static is the most straightforward parallelization strategy as decisions are taken before application execution. 
		The chunk sizes and their assignment to PEs are known before execution.
		The \texttt{STATIC}~scheduling technique produces the smallest scheduling overhead, it favors regular loops with constant-length iterations executing on homogeneous and non-noisy systems. 
		An example of static loop scheduling techniques is Static Chunking or simply \texttt{STATIC}. \mbox{Eq.~\ref{eq:static}} shows how the chunks are calculated, a loop is decomposed into $P$ equal-sized chunks of iterations and each PE receives one of those chunks in order. 
		That is, the chunk$_\text{zero}$ goes to PE$_\text{zero}$, the chunk$_\text{one}$ goes to PE$_\text{one}$, etc.
	}

	\begin{equation}\label{eq:static}
		Cs = \frac{N}{P}
	\end{equation}
	
	\hl{
		Apart from inherent load imbalance in application kernels and system variability, one of the motivations for considering dynamic scheduling techniques is to address the asynchronous arrival of threads at the start of loop execution, which can be due to operating system noise, NUMA placement, etc. 
		When threads begin processing at different times, any dynamic scheduling strategy handles the resulting imbalance by self-assigning work to threads as they become available. 
		Threads that arrive after all iterations have been consumed will not self-assign any work and simply wait and/or die at the implicit barrier of parallel \textit{for} regions.
		In contrast, static scheduling schemes typically assume synchronized thread start times, which can lead to idle threads and inefficient execution in the presence of delays. 
		Each scheduling algorithm considered in this work (and detailed in the following) is implemented with an execution control mechanism that uses mutexes and/or atomic operations to coordinate asynchronous thread access to the central work queue where iterations are stored in the OpenMP runtime. 
		These synchronization constructs are specifically designed to ensure correct and efficient behavior when threads compete asynchronously for the next chunk of iterations, which is most often the case.
	}

	\hl{
		In this work, we consider the following \textbf{dynamic and non-adaptive self-scheduling} algorithms, which follow a predefined function to calculate chunk sizes based on the number of workers (threads) and the number of remaining loop iterations.
	}
	\begin{itemize}
		\item \hl{self-scheduling \mbox{(SS)~\cite{SS}}};
		\item \hl{guided self-scheduling \mbox{(\texttt{GSS})~\cite{GSS:1987}}};
		\item \hl{trapezoid self-scheduling \mbox{(\texttt{TSS})~\cite{TSS:1993}}};
		\item \hl{\mbox{factoring~(\texttt{FAC})~\cite{FAC:1992}}};
		\item \hl{a practical variant of \mbox{factoring~(\texttt{FAC2})}};
	\end{itemize}

	\hl{
		With \mbox{\texttt{SS}~\cite{SS}}, the size of the chunks is always equal to one iteration; see \mbox{Eq.~\ref{eq:ss}}. 
		This technique often incurs the highest scheduling overhead, as the number of scheduling rounds is the highest, always equal to $N$. 
		The increased number of scheduling rounds can increase locality issues, the time spent on synchronization procedures, and the time spent on the assignment itself.
		On the other hand, \texttt{SS} achieves the highest load balance, as PEs always receive the smallest possible unit of work (one iteration in the context of loop scheduling).
	}
	
	\begin{equation}\label{eq:ss}
		Cs = 1
	\end{equation}
	
	\hl{
		With \mbox{\texttt{GSS}~\cite{GSS:1987}} PEs obtain chunks of decreasing sizes instead of a fixed size. 
		The first chunks are large in size, decreasing at every scheduling round until one iteration per chunk is reached.
		\texttt{GSS} aims to reduce the scheduling overhead of \texttt{SS}~with fewer chunks and still achieve load balance by scheduling small chunks of iterations towards the end of the loop execution. 
		\mbox{In~\cite{GSS:1987}}, the authors assume that PEs have unequal starting times caused by, for instance, other work prior to the loop or different thread wake-up times. 
		Motivated by this problem, \texttt{GSS}~was designed with decreasing chunk sizes following \mbox{Eq.~\ref{eq:gss}}. 
	}
	
	\begin{equation}\label{eq:gss}
		Cs_{i}=\left\lceil\frac{R_{i}}{P}\right\rceil
	\end{equation}
	
	\hl{
		\mbox{\texttt{TSS}~\cite{TSS:1993}} follows the same idea as \texttt{GSS}~with a slightly different function that decreases the chunk sizes linearly. 
		\texttt{TSS}~can take two inputs from the user which specify the size of the first chunk, $f$, and last chunk $l$. 
		Then, the chunk size is calculated following \mbox{Eq.~\ref{eq:tss}}.
		As a general suggestion, the authors \mbox{of~\cite{TSS:1993}} recommend the use of the first chunk size $f=\frac{N}{2P}$. 
	}
	
	\begin{equation}
		\label{eq:tss}
		\begin{aligned}
			A&=\left\lceil\frac{2N}{f+l}\right\rceil,\\
			\delta&=\frac{f-l}{A-1},\\
			Cs(1)&=f,\\
			Cs_{i}&=(Cs_{i}-1)-\delta
		\end{aligned}
	\end{equation}
	
	\hl{
		\mbox{\texttt{FAC}~\cite{FAC:1992}} aims to better address the variance in iteration execution times than \texttt{GSS} and \texttt{TSS}.
		\texttt{FAC} evolved from comprehensive probabilistic analyses, and assumes prior knowledge about the average iteration execution times ($\mu$) and their standard deviation ($\sigma$). 
		In contrast to earlier methods, this technique schedules iterations in batches of $P$ chunks of equal size. 
		For each batch, a chunk size is calculated according to \mbox{Eq.~\ref{eq:fac}} and then $P$ chunks of the calculated size are created and scheduled in the following scheduling rounds until the batch ends. 
		\texttt{FAC}~becomes similar to \texttt{GSS} when each batch contains only one chunk, and similar to FSC when only one batch is created.
	}
	
	\begin{equation}\label{eq:fac}
		\begin{aligned}
			Cs_{j}&=\left\lceil\frac{R_{j}}{x_{j}P}\right\rceil,\\
			R_{0}&=N, R_{j+1}=R_{j}-PCs_{j},\\
			b_{j}&=\frac{P}{2\sqrt{R_{j}}}\times\frac{\sigma}{\mu},\\
			x_{0}&=1+b_{0}^{2}+b_{0}\sqrt{b_{0}^{2}+2},\\
			x_{j}&=2+b_{j}^{2}+b_{j}\sqrt{b_{j}^{2}+4}, j>0
		\end{aligned}
	\end{equation}
	
	\hl{
		\mbox{\texttt{FAC2}~is} a simplified and more practical version of \texttt{FAC}. 
		\mbox{\texttt{FAC2}~sets} $x=2$, eliminating the need for $\sigma$ and $\mu$.
		Another more efficient practical version of \texttt{FAC2} is \texttt{mFAC2}, proposed \mbox{in~\cite{lb4omptpds}} and considered in this work.
		\texttt{mFAC2} follows the same pattern as \texttt{FAC2} differing only in implementation for lower overhead. 
		While \texttt{FAC2} uses a mutex to compute and share the chunk size for each batch, introducing synchronization overhead, \texttt{mFAC2} avoids this by using an atomic counter to identify the current batch. 
		Each thread then computes its own chunk size based on this counter. The choice between \texttt{FAC2} and \texttt{mFAC2} depends on the relative costs of synchronization and computation on the target system. 
		In this work, we use \texttt{mFAC2} as previous \mbox{works~\cite{lb4omptpds, auto4omp}} show that it outperforms the default \texttt{FAC2} implementation.
	}
	
\begin{figure}[!htb]
		\centering
		\includegraphics[width=0.6\columnwidth]{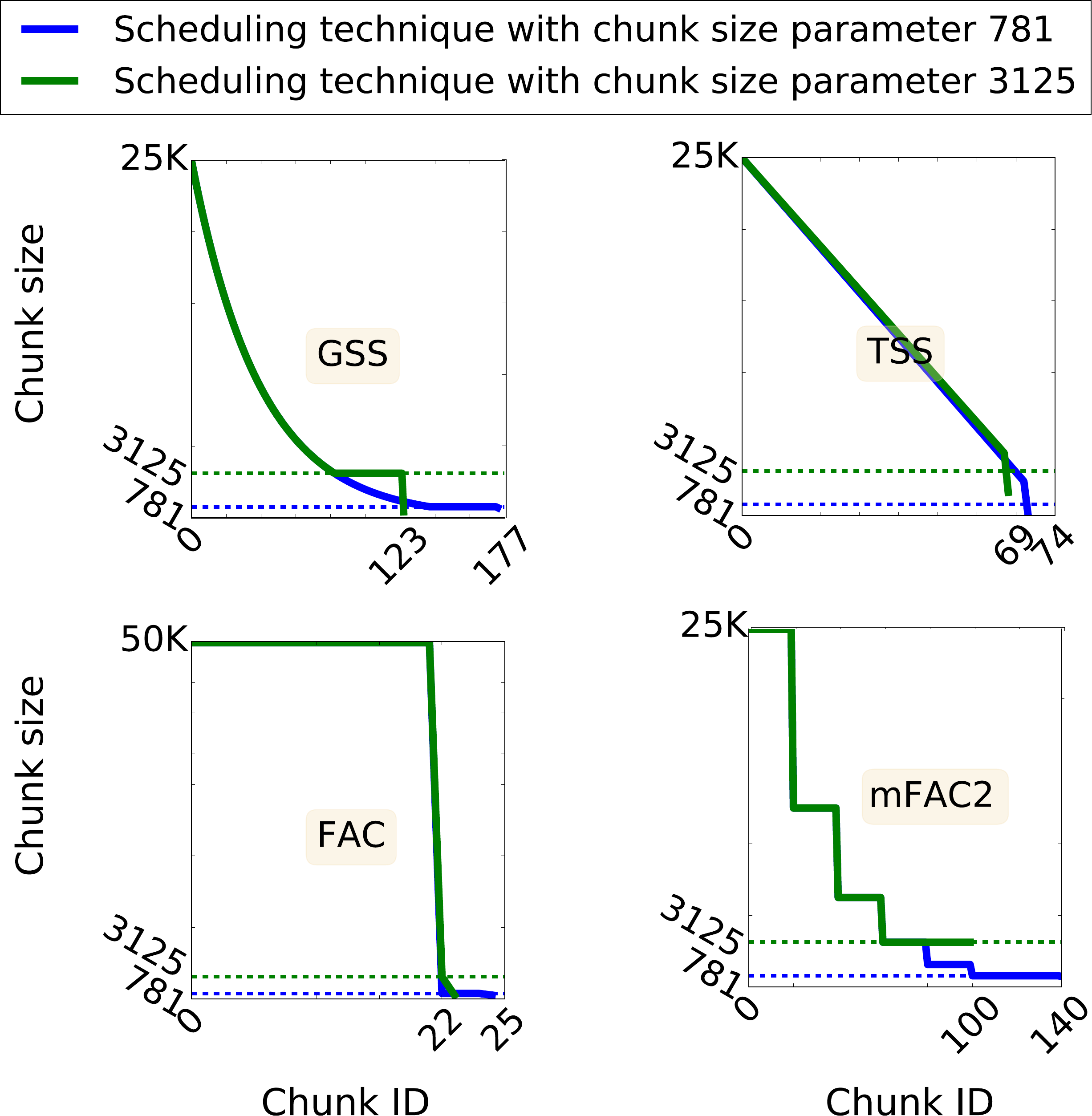}
		\caption{
			\hl{
				Progression of chunk sizes for \textbf{dynamic} and \textbf{non-adaptive} techniques for scheduling the most time consuming loop (gravity calculation across particles) of \sphynx with $1,000,000$ loop iterations on a 20-thread \xeon node and two chunk parameters of 781 and 3125 loop iterations. 
				The two chunk parameter were kept here to highlight their impact on the progression of chunk sizes calculated by the different scheduling algorithms.
				Chunk IDs are shown on the $x$ axis and their sizes on the $y$ axis.
				Figure adapted \mbox{from~\cite{lb4omptpds}}.
			}
		}
		\label{fig:candynonadapt}
	\end{figure}

	\hl{
		\mbox{Fig.~\ref{fig:candynonadapt}} illustrates how chunk sizes evolve across work requests for the \textit{dynamic} and \textit{non-adaptive} scheduling algorithms when scheduling the main loop (\textbf{L1}) of \mbox{\sphynx~\cite{sphynx}} (see \mbox{Section~\ref{sec:factable}} for more details).
		\mbox{Fig.~\ref{fig:candynonadapt}}, since their chunk sizes remain constant (represented as a straight line), defined by the chosen \textit{chunk parameter}.
		The \textit{dynamic} \mbox{\textit{non-adaptive}} strategies exhibit a trend of decreasing chunk size, where each subsequent chunk is less than or equal to the previous one.
		It is evident that both the size of the chunks and the total number of chunks allocated differ between the techniques.
		Fewer and larger chunks may result in less scheduling overhead but a greater risk of imbalance, while a larger number of smaller chunks can enhance load balance at the cost of increased scheduling operations.
	}

	\hl{
		We also consider the following \textbf{dynamic and adaptive scheduling techniques} that adjust chunk sizes during execution based on runtime measurements.
	}
	\begin{itemize}
		\item \hl{adaptive weighted factoring variants (\texttt{AWF-B}, \texttt{AWF-C}, \texttt{AWF-D}, \mbox{\texttt{AWF-E})~\cite{AWF:2003};}}
		\item \hl{adaptive factoring \mbox{(\texttt{AF})~\cite{AF:2000}}}.
	\end{itemize}

	\hl{
		The \mbox{\texttt{AWF}~\cite{AWF:2003}} variants are capable of updating the weights of the PEs during the execution of a loop to calculate how many iterations each thread should receive on every scheduling round.
		\texttt{AWF-B} schedules the iterations in batches and updates the weights after each batch based on timings from previous batches.
		\texttt{AWF-C} schedules the iterations in chunks instead of batches. 
		The idea of this variation is to address a possible issue of \texttt{AWF-B} (as well as \texttt{FAC}, \texttt{WF} and \texttt{AWF}), where faster PEs that already computed their portion of the batch could be assigned remaining chunks of less-than-optimal size from the current batch. 
		\texttt{AWF-C} recomputes a new batch each time a PE requests work. 
		This makes faster PEs get larger chunks from all the remaining iterations and not just from the ones left in the current batch.
		\texttt{AWF-D} is similar to \texttt{AWF-B} but the execution time of iterations of a chunk $j$, is redefined as the total chunk time.
		\texttt{AWF-E} is similar to \texttt{AWF-C} but uses to total chunk time as in \texttt{AWF-D} to recalculate weights.
	}
	
	\hl{
		\mbox{\texttt{AF}~\cite{AF:2000}} is based on \texttt{FAC} and relaxes the requirement of \texttt{FAC} for the $\mu$ and $\sigma$ to be known before the execution. 
		The values of $\mu$ and $\sigma$ are calculated and updated during execution. 
		As \texttt{FAC}, \texttt{AF} uses a probabilistic model to calculate the chunk sizes. 
		In comparison to \texttt{WF}, \texttt{AF} does not use fixed weights but dynamically calculates the size of a new chunk for a PE based on its performance information gathered from recently executed chunks. 
		The chunk calculation of \texttt{AF} is shown in \mbox{Eqs.~\ref{eq:af-c}} and \mbox{\ref{eq:af-dt}}.
	}
	\begin{equation}
		\label{eq:af-c}
		\scalebox{0.9}{%
			$\begin{aligned}
				&Cs_{i}^{\left(n\right)}=\frac{D_{n}+2T_{n}R_{n-1}-\sqrt{D_{n}^{2}+4D_{n}T_{n}R_{n-1}}}{2\hat{\mu}_{i}},\quad n>1 \\
				&\text{and}\quad Cs^{\left(1\right)}\geq 100,\quad n=1
			\end{aligned}$
		}
	\end{equation}
	
	\begin{equation}
		\label{eq:af-dt}
		\scalebox{0.9}{%
			$\begin{aligned}
				&D_{n}=\sum_{i=1}^{P}{\frac{{\sigma}_{i}^{2}}{{\mu}_{i}}}, \quad
				T_{n}=\left(\sum_{i=1}^{P}{\frac{1}{{\mu}_{i}}}\right)^{-1}
			\end{aligned}$
		}
	\end{equation}
	
\begin{figure}[!htb]
		\centering
		\includegraphics[width=0.6\columnwidth]{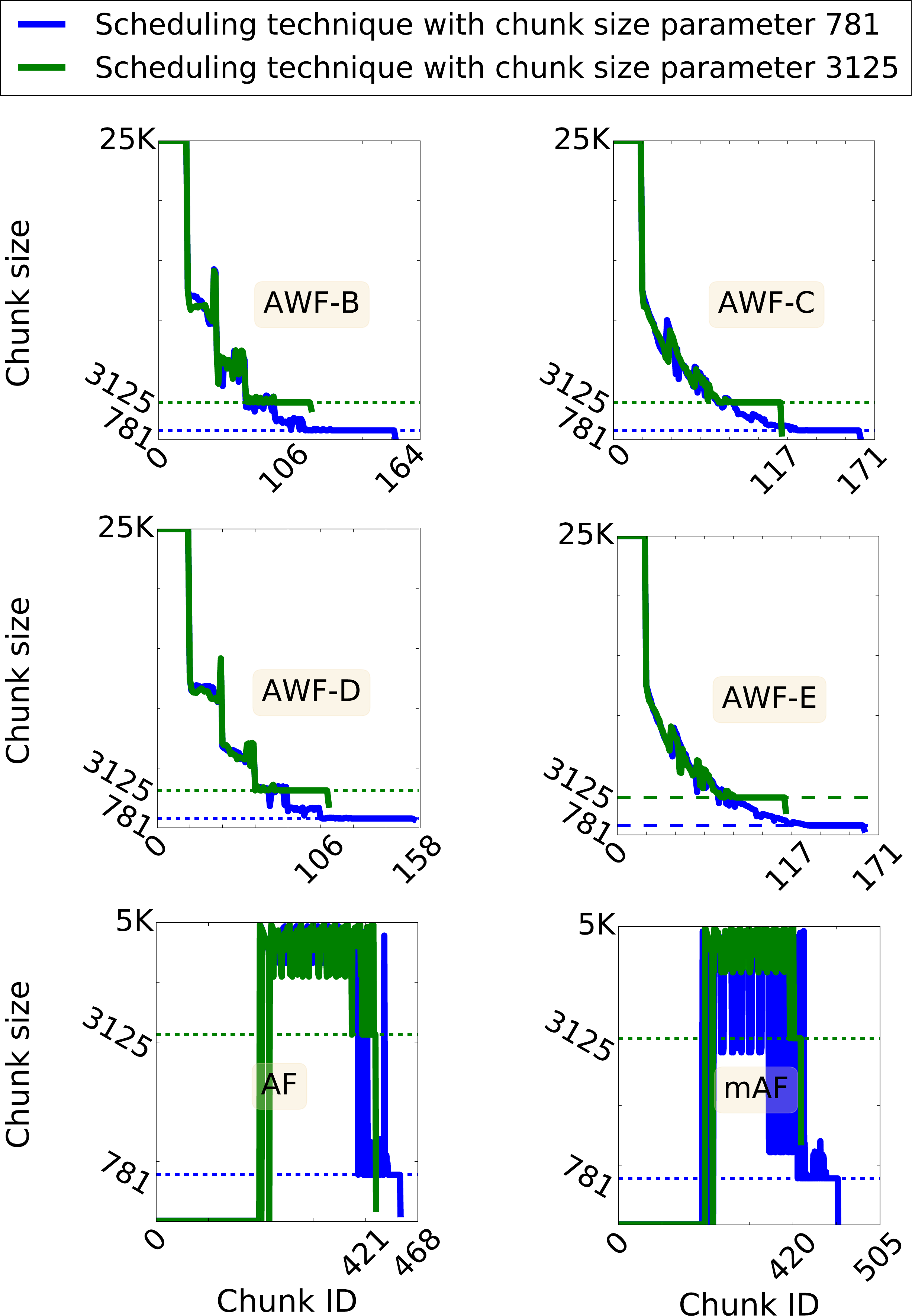}
		\caption{
			\hl{
				Progression of chunk sizes for \textbf{dynamic} and \textbf{adaptive} techniques for scheduling the most time consuming loop (gravity calculation across particles) of SPHYNX with $1,000,000$ loop iterations and 20 threads executed on an Intel Broadwell E5-2640 v4 (2 sockets, 10 cores each) CPU node.
				The chunk parameters here were also 781 and 3125 loop iterations. 
				That is, any scheduling technique will not assign fewer than 781 or 3125 iterations per work request.
				Chunk IDs are shown on the $x$ axis and their sizes on the $y$ axis. 
				\texttt{mAF} and \mbox{\texttt{AF}~only} differ in terms of practical implementation, as described in earlier \mbox{work~\cite{lb4omptpds}}. 
				Figure adapted \mbox{from~\cite{lb4omptpds}}.
			}
		}
		\label{fig:candyadapt}
	\end{figure}

	\hl{
		\mbox{Figure~\ref{fig:candyadapt}} shows the chunk sizes calculated by the dynamic and \textbf{adaptive} scheduling techniques and their progression over the work requests for the main loop of the SPHYNX \mbox{application~\cite{sphynx}}. 
		The chunk sizes calculated by the adaptive scheduling techniques do not strictly decrease with each work request but increase or decrease depending on the requesting thread's or process's performance during execution.
		\textbf{This is the quintessence of adaptive self-scheduling and load balancing: workers that require more time to compute receive less work while workers that compute faster receive more work.}
	}
	
	\section{Automated Scheduling Algorithm Selection}\label{sec:main}
	
	
	\hl{Selecting an appropriate scheduling algorithm in OpenMP can have a substantial impact on application performance, particularly for loops with irregular or evolving workloads. 
		This section presents the automated scheduling algorithm selection methods approached in this work. }
	
	\hl{These methods were designed to adapt, being capable of selecting and switching between different scheduling algorithms during runtime. 
		We focus on two categories: expert-based and reinforcement learning (RL)-based methods. }
	\hl{While the expert-based approaches build on our prior work}~\cite{auto4omp}\hl{, this paper introduces and integrates RL-based methods into the OpenMP runtime, more specifically using the LB4OMP library}~\cite{lb4omptpds}. 
	\hl{We focus on comparing these strategies to assess their effectiveness, trade-offs, and potential for improvement.}
	
	\subsection{Scheduling Algorithms Portfolio in OpenMP}\label{sec:algo-portfolio}
	
	The expert- and RL-based methods considered in this work (described in Sect.~\ref{sec:expertsel} and~\ref{sec:rlsel}) select from a portfolio of 12 scheduling algorithms:
	\begin{enumerate}[align=left, leftmargin=*]
		\item The three scheduling algorithms specified in the OpenMP standard: \static, Dynamic (\texttt{SS})~\cite{SS}, and Guided (\gss)~\cite{GSS:1987}.
		\item Three scheduling algorithms that were already available in the LLVM OpenMP runtime library prior to the LB4OMP release: Trapezoid self-scheduling (\tss)~\cite{TSS:1993}, \steal~\cite{worksteal99}, and \texttt{schedule(auto)} available in LLVM (\llvmauto).
		\item Six LB4OMP scheduling algorithms~\cite{lb4omptpds}: the practical variant of Factoring (\factwo)~\cite{FAC:1992}, four variants of Adaptive Weighted Factoring (\awfb, \awfc, \awfd, and \awfe)~\cite{AWF:2003}, and Adaptive Factoring (\af)~\cite{AF:2000}.
	\end{enumerate}
	
	
	\hl{
		All selection methods and scheduling algorithms (both RL-based and expert-based) are designed to operate at the loop level and can be applied independently to multiple loops within the same application. 
		\mbox{\lbomp~\cite{lb4omptpds}} assigns a unique identifier to each loop instance, allowing the selection mechanisms to track and adapt to the behavior of each loop individually as / if / when it reappears. 
		For example, in the \mandel{} application, the selection framework independently instrumented and handled three separate loops (see \mbox{Sect.~\ref{sec:perfanalysis})}.
	}
	
	This portfolio, previously used in the literature~\cite{auto4omp}, is considered here to facilitate a fair comparison between expert- and RL-based methods. Additionally, it allows us to assess whether the portfolio designed for expert-based methods is also advantageous for RL-based approaches.

	\subsection{Expert-based Scheduling Algorithm Selection}\label{sec:expertsel}
	
	The expert-based automated scheduling algorithm selection methods used in this work, \autoRND, \autoEXT, and \autoEXP,  were proposed in~\cite{auto4omp}. 
	They incorporate expert knowledge of the execution performance of the LB4OMP algorithms across various applications and systems.
	
	\hl{
		\mbox{\textbf{\autoRND}~\cite{auto4omp}} 
		selects a scheduling algorithm at random, without explicitly searching for the best-performing option. 
		It introduces the concept of a jump probability, \mbox{$P_j$}, which determines the likelihood of changing the current scheduling algorithm within a predefined \mbox{range—starting} from \mbox{DLS$_0$} = \texttt{STATIC} up to \mbox{DLS$_n$} = \texttt{mAF}.
	}
	
	\hl{
		At each loop execution instance (time-step), \mbox{\autoRND{}} generates a random number $RND$ uniformly distributed between 0 and 1. If $P_j > RND$, a new scheduling algorithm is randomly selected from the \mbox{\Auto}~portfolio. Otherwise, the previously selected algorithm is retained.
	}
	
	\hl{
		The value of $P_j$ is defined as $LIB/10$, where $LIB$ is the percent load imbalance \mbox{metric~\cite{derose2007detecting}}, calculated as shown in \mbox{\Eq~\ref{eq:pi}}. The denominator (10) is an empirically chosen constant. This design ensures that when the load imbalance exceeds $10\%$, \mbox{\autoRND{}} will always select a new scheduling algorithm. The threshold of $LIB = 10$ has been empirically identified as a practical indicator of high load \mbox{imbalance—situations} where the potential benefits of switching algorithms are likely to outweigh the associated overhead.
	}
	\begin{equation}
		\label{eq:pi}
		\begin{split}
			&LIB=\left( 1- \frac{\text{\texttt{mean} of thread finishing times}}{\text{\texttt{max} of thread finishing times}} \right) \times 100.
		\end{split}
	\end{equation}
	
	\hl{
		\mbox{\autoEXT~\cite{auto4omp}} performs an exhaustive search by evaluating every scheduling algorithm in the \mbox{\Auto}~\mbox{portfolio—ranging} from DLS$_0$ = \texttt{STATIC} to DLS$_n$ = \mbox{\texttt{mAF}—over} successive loop execution instances (time-steps). 
		At each time-step, a different algorithm is applied, and its execution time is recorded. 
		After all algorithms have been tested once, the algorithm with the shortest observed execution time is selected for continued use. 
		This approach guarantees a comprehensive evaluation of available options while incurring a trial overhead proportional to the number of algorithms in the portfolio.
	}
	
	\hl{
		Rather than repeating the same loop time-step with different algorithms, \autoEXT{} changes the scheduling algorithm on each new execution of the loop, progressively advancing the application's execution while accumulating performance data. 
		The scheduling decision is then based on these recorded execution times.
		Since \mbox{\autoEXT{}} compares execution times across different time-steps, e.g., using \texttt{STATIC} at time-step 0 and \texttt{SS} at time-step 1, its decisions rely on consistent loop behavior across time-steps. 
		If execution time varies significantly due to dynamic workload changes or system noise, this may lead to sub-optimal algorithm selection.
	}
	
	\hl{
		Once the highest-performing algorithm has been identified, it is used for scheduling the remaining loop iterations. 
		Meanwhile, the load imbalance metric $LIB$ (see \mbox{\Eq~\ref{eq:pi}}) continues to be computed after each loop execution. 
		If a high imbalance is detected ($10\%$ variation from the recorded average) with the selected algorithm, the exhaustive search is re-triggered to reassess and select a more suitable scheduling strategy.
	}
	
	\hl{
		\mbox{\autoEXP{}~\cite{auto4omp}} selects a suitable scheduling algorithm at each loop execution instance, similarly to \autoRND{}, yet it relies on expert knowledge and fuzzy \mbox{logic~\cite{zadeh1996fuzzy}} rather than randomness. 
		It makes decisions based on performance data collected during \mbox{execution—specifically}, loop execution time ($T_{par}$) and load imbalance ($LIB$).
	}

	\hl{
		The first loop instance is executed with \texttt{STATIC} to collect initial $T_{par}$ and $LIB$ values. 
		From the second execution onward, \autoEXP{} uses two fuzzy logic systems: one for the initial algorithm selection based on absolute values, and another for subsequent adjustments based on changes in $T_{par}$ and $LIB$ over time (\mbox{in~\cite{auto4omp}}, see Sect. 3.3.3, Fig. 5, Tab. 1, and Algorithm 1).
	}
	
	\hl{
		Fuzzy logic enables reasoning under uncertainty by mapping numeric inputs to qualitative categories (e.g., "Low", "Moderate") and applying expert rules to guide decisions. 
		This approach simplifies the encoding of scheduling expertise and enables adaptive, rule-based algorithm selection without the need for exhaustive trials or random jumps.
	}


	These selection methods also benefit from an expert chunk parameter 
	(\mbox{\expchunk}).
	\texttt{chunk} parameter has been introduced in the OpenMP standard to minimize scheduling overhead and improve data locality.
	The \expchunk{} is calculated using the golden ratio $\phi=1.618$, the number of loop iterations $N$, and the number of threads $P$ in the interval $[N/(2P),...,1]$ to find a chunk parameter value that leads to high performance, using only $N$, $P$ and $\phi$ (see~\cite{auto4omp} in Sect. 3.1, Eq. 1).
	In~\cite{auto4omp}, they show that the \expchunk{} parameter increases the performance of all scheduling algorithms in most cases.

	\subsection{Oracle Scheduling Algorithm Selection}\label{subsub:oracle}
	
	The \gt{} represents a \textit{manually} selected set of scheduling algorithms and chunk parameters (default or \expchunk) optimized for peak performance across all combinations of application, loop, time-step, and system. 
	It is derived from experiments with each of the 12 algorithms in the portfolio. 
	
	Although \gt{} may not be practically achievable, \textit{it serves as a theoretical upper performance limit for application loops}, helping us to assess the proximity of expert- and RL-based methods to this ideal as a baseline. 
	
	It is important to recognize that the performance indicated by \gt{} is specific to the scenarios tested and does not account for system variability or the inherent nondeterminism in parallel application executions.

	\subsection{Reinforcement Learning Selection Methods}\label{sec:rlsel}
	
	Reinforcement Learning (RL) is a subset of machine learning (ML)~\cite{zhou2021machine} where agents interact with an environment to make decisions based on the rewards received~\cite{suttonReinforcement1998}. 
	RL methods are classified as \emph{model-based} or \emph{model-free}.
	\emph{Model-based RL} approaches such as I2A~\cite{i2a2017} and Prioritized Sweeping~\cite{corneil2018efficient}, derive a model to predict possible rewards for actions. 
	\emph{Model-free RL} approaches, on the other hand, use estimated rewards without an explicit model of the environment~\cite{ramirez2022model}.
	
	\emph{We use model-free RL approaches}, specifically \sarsa{} and \qlearn, due to the lack of a suitable model and the absence of extensive data required to design one. 
	In dynamic environments such as HPC applications and systems, scheduling algorithm selection poses challenges that preclude the use of model-based methods. 
	\emph{This study serves as an initial step towards leveraging the data collected here to define such a model-based RL.}

	\sarsa{} is a learning strategy where the state$\rightarrow$action value $Q(s, a)$ is updated based on the received reward $r$ and new state$\rightarrow$action value $Q(s^{'}, a^{'})$. In Eq.~\ref{eq:sarsa}, $s$ and $a$ are the current state and action, while $s^{'}$ and $a^{'}$ are the next state and action. The learning rate $\alpha$ (0 to 1) dictates the update rate of $Q(s, a)$, with higher $\alpha$ leading to faster updates. The discount factor $\gamma$ (0 to 1) determines the influence of future rewards, with higher values giving more weight to future rewards.
	\begin{equation} \label{eq:sarsa}
		Q(s, a) = Q(s, a) + \alpha \left( r + \gamma \times Q(s^{'}, a^{'}) - Q(s, a) \right)
	\end{equation}
	
	\qlearn{} is similar to \sarsa{}, as shown in Eq.~\ref{eq:qlearn}. 
	The greedy action strategy differs from \sarsa's as \qlearn{} greedily chooses the action that maximizes the state$\rightarrow$action value. This is indicated by the term $\max_{a^{'}}\left(Q(s^{'}, a^{'})\right)$ below.
	\begin{equation} 
		\label{eq:qlearn}
		Q(s, a) = Q(s, a) +
		\alpha  \left( r +  \gamma \times \max_{a^{'}} \left(Q(s^{'}, a^{'})\right) - Q(s, a) \right)
	\end{equation}

	\subsection{RL-based Scheduling Algorithm Selection}\label{sec:RLsel-in-LB4OMP}

\begin{figure}[!htb]
		\centering
		
		\includegraphics[width=0.8\linewidth]{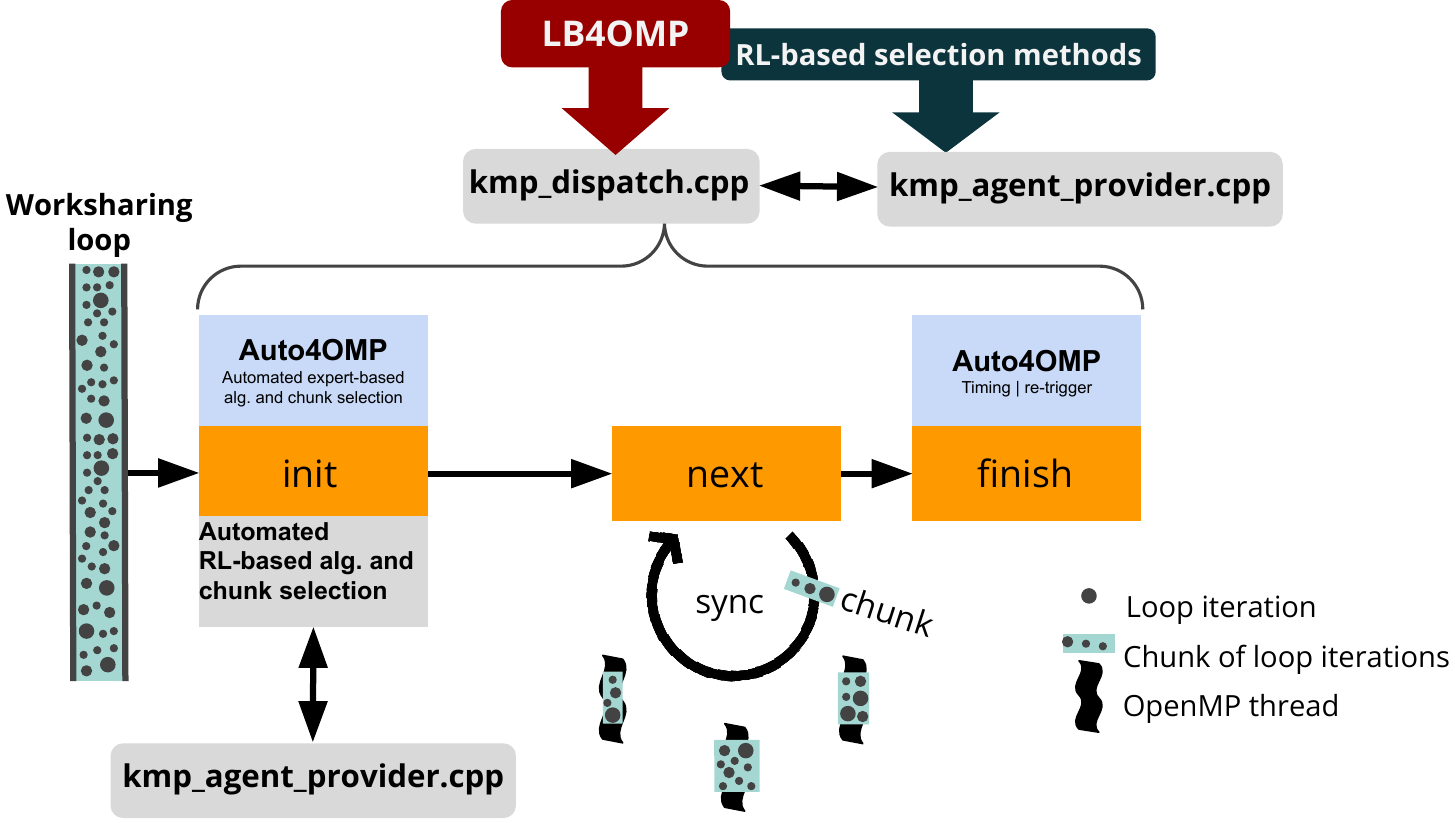}
		\caption{
			Scheduling in \lbomp augmented with RL-based selection of scheduling algorithms.
		}
		\label{fig:rl4ompsched}
		
	\end{figure}

	The RL-based methods were implemented in LB4OMP as a separate component, \texttt{kmp\_agent\_provider.cpp}, which is included by \texttt{kmp\_dispatch.cpp}, the main file for \textit{for loop} scheduling in the LLVM OpenMP RTL library and LB4OMP (see Fig.~\ref{fig:rl4ompsched}). 
	\hl{This separation allows \textit{the RL methods to be compiled independently and reused in other libraries with minimal modifications, given the appropriate parameters.}}
	
	To use RL-based selection methods, one needs to add the \texttt{SCHEDULE(runtime)} clause to the target OpenMP loops and set the \texttt{OMP\_SCHEDULE} environment variable. 
	\hl{Every loop that contains the SCHEDULE(runtime) is handled separately by the library (each loop receives a unique id) such that different loops in the applications may select different scheduling algorithms.}
	Regarding \Auto~\cite{auto4omp}, we extended the OpenMP \textit{auto} implementation with additional options to trigger the chosen RL-based selection method.
	To choose \qlearn, it is necessary to set \texttt{OMP\_SCHEDULE} to ``\texttt{auto,8}'', and \sarsa{} to ``\texttt{auto,10}''. 
	

	The RL-based methods are executed from the init function (Fig.~\ref{fig:rl4ompsched}) inside \texttt{kmp\_dispatch.cpp}.
	This function is executed once before every loop execution. 
	Selection relies on applications that execute the same loop multiple times (e.g., time-stepping applications). 
	All developed methods require a learning phase before selecting an algorithm.

	\textit{Reward types}: We propose two types of rewards: loop time (\texttt{LT}) and \texttt{LIB}. The \texttt{LT} reward gives positive feedback for faster executions, negative for slower executions, and neutral for comparable times. The \texttt{LIB} reward uses the \textit{execution imbalance percentage}~\cite{derose2007detecting} to evaluate the imbalance of the loop instance, giving positive rewards for low imbalance, negative for high and neutral for comparable values.
	
	The reward function follows Eq.~\ref{eq:reward-function}. 
	The input $x$ is defined depending on the reward type, e.g., \texttt{LIB} denoting \textit{imbalance percentage}. 
	To distinguish between these positive, negative, or neutral rewards, we record $\min$ and $\max$ for the input $x$ across all loop instances already executed.
	\begin{equation}
		\mathcal{R}_t(x) = 
		\begin{cases} 
			r_+ & x \leq \min\limits_{t} (x) \\
			r_0 & \min\limits_{t}(x) < x < \max\limits_{t}(x) \\
			r_- & \max\limits_{t}(x) \leq x) 
		\end{cases}
		\label{eq:reward-function}
	\end{equation}  
	
	The Q-value tables are initialized with 0 for all possible state$\rightarrow$actions -- 144 combinations of scheduling algorithms.
	We consider 0.01 for positive rewards, -4.0 for negative, and -2.0 for neutral. 
	We use 0.01 as a positive reward value, instead of the commonly used 0, to account for a corner case in which the agent would not distinguish between a high-performing action rewarded with 0 and the initial Q-value table value of 0.
	
	\textit{Agent's policy:} We implemented the \texttt{explore-first} policy as the learning phase strategy of RL-based selection methods. 
	This policy selects every scheduling algorithm from the portfolio considering all possible different orders at the beginning of the learning process. 
	This means that the exploration phase requires 144 loop instances to be completed.

	\textit{General configurations and additional functionalities:} 
	The default values for $\alpha$ and $\gamma$ are 0.5.
	Both $\alpha$ and $\gamma$ can, respectively, be configured by setting two different environment variables, \texttt{KMP\_RL\_ALPHA} and \texttt{KMP\_RL\_GAMMA}.
	We also implemented the concept of \textit{alpha decay}, which lowers the value of $\alpha$ (the learning rate) in every loop instance after the learning phase to prevent agents from ever selecting a single action. 
	The decay of $\alpha$ is set to 0.05 and can be configured with an environment variable called \texttt{KMP\_RL\_ALPHA\_DECAY}.
	
	For debugging purposes and to collect additional performance data, we developed a functionality that stores the Q-value tables after each loop instance. 
	\hl{This functionality can be activated with the environment variable \texttt{KMP\_RL\_AGENT\_STATS} which requires a location path to store the data.  
		This can be extended in the future and used to initialize the Q-value tables of applications that have already been executed on a given system. 
		Thus, eliminating the learning phase of RL-based methods.}
	
	All configurations and additional functionalities can be configured through different environment variables. 
	Detailed information on them is available open source at~\cite{lb4ompRepo}.


	\subsection{Research Questions} 
	As we are interested in \textit{learning how to select scheduling algorithms in OpenMP}, we formulate three research questions. 
	
	
	\begin{enumerate}[align=left, leftmargin=*]
		\item[RQ1] How do expert- and RL-based selection approaches compare?
		\item[RQ2] Which parameters influence RL-based selection and in what way?
		\item[RQ3] How can expert- and RL-based selection approaches be combined to harness their joint potential? 
	\end{enumerate}
	
	\section{Experiments and Analysis}\label{sec:perfanalysis}
	
	We use the following notation to describe certain features of the application and system.
	$T$ is the number of \mbox{time-steps}, $P$ the number of threads, $N$ the number of iterations, and $L$ the IDs of loops with modified \texttt{schedule} clauses. 

\subsection{Design of Factorial Experiments}\label{sec:factable}

	For performance evaluation and comparison of expert- and RL-based selection methods, we conducted a total of 3,600 experiments, following the factorial design summarized in Tab.~\ref{table:exprl4omp}. 
	All codes and LB4OMP were compiled with Intel compiler version 2021.6.0.

	\begin{table}[!htb]
		
		\centering
		\caption{Design of factorial experiments, resulting in 720 different configurations and a total of 3'600 executions.}\label{table:exprl4omp}
        \makebox[\textwidth][c]{
		\resizebox{1\textwidth}{!}{
			\begin{threeparttable}
				
				\begin{tabular}{l|l|l|l|l}
					\hline
					\multicolumn{2}{l|}{\textbf{Factors}} &
					\multicolumn{2}{l|}{\textbf{Values}} &
					\textbf{Properties} \\ \hline
					\multicolumn{2}{l|}{\multirow{16}{*}{\textbf{Applications}}} &
					\multicolumn{2}{l|}{\mandel} &
					\begin{tabular}[c]{@{}l@{}}\textbf{Compute-bound kernel} \\ $N$ = $262,144$ iterations$|$ $T$ = $500$ time-steps $|$ Total loops = $3$ $|$ Modified loops = $3$ \\ L0 = constant workload imbalance, \\L1 = increasing workload imbalance, \\L3 = decreasing workload imbalance \end{tabular}\\ \cline{3-5}
					\multicolumn{2}{l|}{\multirow{1}{*}{}} &
					\multicolumn{2}{l|}{\streamtriad} &
					\begin{tabular}[c]{@{}l@{}}\textbf{Memory-bound Benchmark} \\ $N$ = $2,000,000,000$ iterations $|$ $T$ = $500$ time-steps $|$ Total loops  = $1$ $|$ Modified loops = $1$ \\  L0 = no workload imbalance \end{tabular}\\ \cline{3-5}
					\multicolumn{2}{l|}{\multirow{1}{*}{}} &
					\multicolumn{2}{l|}{Triangle Counting (\tc)} &
					\begin{tabular}[c]{@{}l@{}}\textbf{Graph theory and network analysis kernel} \\ $N$ = $1,048,576$ iterations $|$ $T$ = $500$ time-steps $|$ Total loops  = $1$ $|$ Modified loops = $1$ \\ L0 = highly imbalanced due to sparse input\end{tabular}\\ \cline{3-5}
					\multicolumn{2}{l|}{\multirow{1}{*}{}} &
					\multicolumn{2}{l|}{\hacc} &
					\begin{tabular}[c]{@{}l@{}}\textbf{Cosmology N-body code, HACC's particle force kernels} \\ $N$ = $600,000$ iterations $|$ $T$ = $500$ time-steps $|$ Total loops  = $3$ $|$ Modified loops = $1$ \\ L0 = no workload imbalance \end{tabular}\\ \cline{3-5}
					\multicolumn{2}{l|}{\multirow{1}{*}{}} &
					\multicolumn{2}{l|}{\lulesh} &
					\begin{tabular}[c]{@{}l@{}}\textbf{Computational fluid dynamics miniapp} \\ $N$ = $21,952,000$ iterations $|$ $T$ = $500$ time-steps $|$ Total loops  = $39$ $|$ Modified loops = $4$ \\ L0, L1, L2, L3 = mildly load imbalanced \end{tabular}\\ \cline{3-5}
					\multicolumn{2}{l|}{\multirow{1}{*}{}} &
					\multicolumn{2}{l|}{\sphynx} &
					\begin{tabular}[c]{@{}l@{}}\textbf{N-body and hydrodynamics simulation scientific application} \\ $N$ = $1,000,000$ particles (iterations) $|$ $T$ = $500$ time-steps $|$ Total loops  = $37$ $|$ Modified loops = $1$ 
						\\ L0 = variable workload imbalance\end{tabular}\\ 
					\hline
					\multirow{4}{*}{\makecell[l]{\textbf{Loop} \\ \textbf{Scheduling}}}  &  \multirow{2}{*}{\makecell[l]{\textbf{OpenMP standard} \\ {\tiny (\textit{also in \lbomp})}}} & \multicolumn{2}{l|}{\static} & Straightforward parallelization \\ 
					\cline{3-5}
					&  & \multicolumn{2}{l|}{\makecell[l]{\ssdynamic, \gss}} & \multirow{2}{*}{Dynamic and non-adaptive self-scheduling algorithms.} \\ 
					\cline{2-4}
					& \multirow{2}{*}{\textbf{LB4OMP library}~\cite{lb4omptpds}} & \multicolumn{2}{l|}{\makecell[l]{\factwo, \llvmauto (LLVM \texttt{schedule(auto)}), \\ \tss, \steal}} \\ 
					\cline{3-5}
					&& \multicolumn{2}{l|}{\awfb, \awfc, \awfd, \awfe, \af} & Dynamic and adaptive self-scheduling algorithms. \\ 
					\hline
					\multirow{3}{*}{\begin{tabular}[c]{@{}l@{}} \textbf{Scheduling} \\ \textbf{Algorithm} \\ \textbf{Selection}\end{tabular} }&
					\begin{tabular}[c]{@{}l@{}} \textbf{Manual} \end{tabular} &
					Manual Selection & \textcolor{darkgreen}{\gt}\tnote{*} &
					\begin{tabular}[c]{@{}l@{}} 
						Ideal combination of the best choices of scheduling algorithm \\ selection across application loops, and time-steps. \\ Selection based on exhaustive experimentation of all scheduling algorithms.\end{tabular} \\ 
					\cline{2-5} 
					&
					\multirow{3}{*}{\begin{tabular}[c]{@{}l@{}} \textbf{Automated from the} \\ \textbf{LB4OMP library} \end{tabular}} 
					&
					\textbf{\Auto~extension}~\cite{auto4omp} & \begin{tabular}[c]{@{}l@{}}
						\textcolor{orange}{\autoRND}\tnote{*} \\ \textcolor{orange}{\autoEXT}\tnote{*} \\ \textcolor{orange}{\autoEXP}\tnote{*} 
					\end{tabular} &
					\begin{tabular}[c]{@{}l@{}}Expert-based scheduling algorithm selection across application loops and time-steps. \end{tabular} \\ \cline{3-5} 
					&
					&
					\textcolor{blue}{\textbf{\rlomp~extension}} & \begin{tabular}[c]{@{}l@{}}
						\textcolor{blue}{\texttt{Q-Learn}\tnote{*}}  \\ \textcolor{blue}{\texttt{SARSA}\tnote{*}}\\ \end{tabular} &
					\begin{tabular}[c]{@{}l@{}}RL-based scheduling algorithm selection across application loops and time-steps. \end{tabular}\\ 
					\hline 
					\multicolumn{2}{l|}{\multirow{3}{*}{\textbf{Chunk parameter}}} & \multicolumn{2}{l|}{Default} & Chunk size =$N/P$ for static and 1 for all other scheduling algorithms. \\ 
					\cline{3-5}
					\multicolumn{2}{l|}{} & \multicolumn{2}{l|}{\textcolor{orange}{ \expchunk}} & Expert chunk parameter: \thead[l]{``a point at $\frac{1}{Golden~ratio (1.618)} = 0.618$ \\ on the curve between $N/(iP)$ and 1, with $i$ increasing in steps of 2''~\cite{auto4omp}.}\\ \hline 
					\multicolumn{2}{l|}{\textcolor{blue}{\textbf{RL general configuration}}} &
					\multicolumn{2}{l|}{\begin{tabular}[c]{@{}l@{}}\textcolor{blue}{$\alpha$=0.5,} \\ \textcolor{blue}{$\gamma$=0.5,} \\ \textcolor{blue}{reward values: positive 0.01, neutral -2.0, negative -4.0}\end{tabular}} & Variables required to configure the RL selection process.
					\\ 
					\hline
					\multicolumn{2}{l|}{\textcolor{blue}{\textbf{RL initializer}}} &
					\multicolumn{2}{l|}{\textcolor{blue}{\texttt{zero}}} & Starts all possible actions with a reward value of 0.
					\\ 
					\hline
					\multicolumn{2}{l|}{\textcolor{blue}{\textbf{RL exploration policy}}} &
					\multicolumn{2}{l|}{\textcolor{blue}{\texttt{explore-first}}} & \begin{tabular}[c]{@{}l@{}}Selects every combination of scheduling algorithms once at the beginning of the learning process. \\Requires 144 time-steps to learn and select the first scheduling algorithm.
					\end{tabular}
					\\ 
					\hline
					\multicolumn{2}{l|}{\textcolor{blue}{\textbf{RL reward type}}} &
					\multicolumn{2}{l|}{\begin{tabular}[c]{@{}l@{}}\textcolor{blue}{\texttt{LT}} \\ \textcolor{blue}{\texttt{LIB}} \end{tabular}} & \begin{tabular}[c]{@{}l@{}} Loop time reward. \\ Load imbalance reward. \end{tabular}
					\\ 
					\hline
					\multicolumn{2}{l|}{\multirow{5}{*}{\textbf{Computing nodes}}} & \multicolumn{2}{l|}{\xeon} & \begin{tabular}[c]{@{}l@{}}Intel Xeon E5-2640 v4 (2 sockets, 10 cores each)\\ $P$ =  20 without hyperthreading, Pinning: \texttt{OMP\_PLACES} = cores \texttt{OMP\_PROC\_BIND} = close \end{tabular} \\ 
					\cline{3-5}
					\multicolumn{2}{l|}{\multirow{3}{*}{}} & \multicolumn{2}{l|}{\gpu} & \begin{tabular}[c]{@{}l@{}}Intel Xeon Gold 6258R (2 sockets, 28 cores each)\\ $P$ =  56 without hyperthreading, Pinning: \texttt{OMP\_PLACES} = cores \texttt{OMP\_PROC\_BIND} = close\end{tabular} \\ 
					\cline{3-5}
					\multicolumn{2}{l|}{} & \multicolumn{2}{l|}{\multirow{2}{*}{\amd}} & \multirow{2}{*}{\begin{tabular}[c]{@{}l@{}}AMD EPYC 7742 (2 sockets, 64 cores each)\\ $P$ =  128 without hyperthreading, Pinning: \texttt{OMP\_PLACES} = cores \texttt{OMP\_PROC\_BIND} = close
					\end{tabular}} \\
					\multicolumn{2}{l|}{} & \multicolumn{2}{l|}{} & \\ \hline
					\multicolumn{2}{l}{} & \multicolumn{2}{l}{} & \\ \hline
					\multicolumn{2}{l|}{\multirow{3}{*}{\textbf{Metrics}}} & \multicolumn{2}{l|}{\thead[l]{$T_{par}^{loop}$}} & Parallel loop execution time (s). \\ \cline{3-5}
					\multicolumn{2}{l|}{} & \multicolumn{2}{l|}{\thead[l]{$execution\:imbalance = \frac{maximum\:time - average\:time}{maximum\:time} \times \frac{P}{P-1}$}} & Execution imbalance (\%)~\cite{derose2007detecting}. \\ 
					\bottomrule
				\end{tabular}
				\begin{tablenotes}
					\small{  
						\item[*] Selects a scheduling algorithm among: \{\texttt{STATIC}, \texttt{SS}, \texttt{GSS}, \llvmauto, \texttt{TSS}, \texttt{Static Steal}, \texttt{mFAC2}, \texttt{AWF-B}, \texttt{AWF-C}, \texttt{AWF-D}, \texttt{AWF-E}, and \texttt{mAF}\}.
						\item[] Text in \textcolor{orange}{orange} highlights the expert-based selection methods and expert chunk parameter proposed in prior work~\cite{auto4omp}.        
						\item[] The text in \textcolor{blue}{blue} denotes the novel contributions of this work and highlights configurations that affect only RL-based selection methods.}
				\end{tablenotes}
			\end{threeparttable}
		} 
        }

	\end{table}

	We consider six applications with loops that vary in memory- or compute-boundedness to test the algorithm selection methods and uncover potential limitations. 
	\textit{Memory-bound} loops challenge dynamic scheduling due to data locality loss, making exploration phases costly for automated methods. \textit{Compute-bound} loops are less affected by data locality loss but are sensitive to load imbalance, requiring selection methods to balance the load with minimal scheduling overhead.

	\textbf{\mandel} is a compute-bound code that calculates the Mandelbrot set and often shows a high load imbalance due to its irregular workload. 
	We implemented it with three OpenMP-parallelized loops enclosed in an outer time-stepping loop.
	The three different loops were designed to 'zoom' into different regions of the \mandel{} set intentionally to exhibit a constant workload imbalance (L0), increasing (L1) and decreasing (L2) over time. 
	\mandel{} was selected to evaluate the sensitivity of the selection methods to dynamic imbalance patterns, making it representative of irregular scientific codes with evolving loop behavior.
	\hl{In addition, \mandel~showcases the ability of the library to handle multiple different loops at the same time.}
	
	\hl{
		\mbox{\textbf{\streamtriad}\cite{streambench}} is a \mbox{memory-bound benchmark} that performs two loads, one store and one fused multiply-add (FMA) per loop iteration. 
		Its loop (L0) has a highly regular workload and is executed repeatedly in a time-stepping fashion. 
		Although \streamtriad~typically achieves load balance execution even with \static scheduling. \streamtriad{} is a challenging problem for dynamic scheduling algorithms and automated selection methods due to its high sensitivity to scheduling overhead and data locality disruptions. 
		This makes it representative of bandwidth-bound kernels often found in stencils and vectorizable codes. 
		Also, we use \streamtriad{} to highlight 'worst case scenarios' that automated methods need to overcome to avoid extreme overhead.
	}
	
	\hl{
		\mbox{\textbf{Triangle counting (\tc)}\cite{gapBench2015}} counts triangles in a graph and is widely used in graph analysis workloads, including biological and social network analysis. 
		The loop (L0) exhibits a severe workload imbalance with \static scheduling  due to the heterogeneous density of the graph, causing uneven execution times of the iterations. 
		We use the implementation of the GAP benchmark suite of \mbox{\tc~\cite{gapBench2015}}, configured with the ``\texttt{-g 20}'' to generate a Kronecker graph with $2^{20}$ vertices, as specified in the \mbox{Graph500~\cite{graph500}} benchmark. 
		This application models real-world graph workloads where irregular parallelism is a core challenge.
	}
	
	\hl{
		\mbox{\textbf{\hacc}\cite{haccKernels2018}} is a compute-bound benchmark based on the \mbox{short-range} force calculation kernels of the Hybrid Accelerated Cosmology Code \mbox{(HACC)~\cite{haccFull2016}}. 
		We modified it to repeatedly compute the \textit{GravityForceKernel6} 500 times in a \mbox{time-stepping} loop to assess performance under low-variance compute-dominated workloads. 
		\hacc{}'s primary loop (L0) shows very low load imbalance with \static scheduling, making it representative of tightly optimized scientific kernels with uniform iteration costs.
	}
	
	\hl{
		\mbox{\textbf{\lulesh}~\cite{lulesh2013}} is a hydrodynamics mini-application that solves a Sedov blast problem using a Lagrangian formulation. 
		It features a mix of memory- and compute-bound behavior. 
		We focus on four of its most time-consuming OpenMP loops: \textit{CalcFBHourglassForceForElems}, \textit{CalcHourglassControlForElems}, \textit{CalcKinematicsForElems}, and \textit{IntegrateStressForElems}. 
		These loops tend to exhibit mild imbalance, with relatively consistent workload per iteration. 
		\lulesh{} represents multiphysics workloads with moderate parallel irregularities and varying loop characteristics.
		\lulesh{} also highlights the capability of \lbomp{} to handle multiple loops at the same time.
	}
	
	\textbf{SPHYNX}\cite{cabezon2017sphynx} is a particle physics application that uses the smoothed particle hydrodynamics (SPH) method to simulate astrophysical phenomena. 
	We ran an Evrard collapse simulation and focused on the \textit{gravity} calculation loop (L0), which is the most time consuming of the 37 OpenMP-parallelized loops in the application. 
	This loop alone accounts for more than $80\%$ of the total execution time and exhibits a variable load imbalance between time steps under \static scheduling\cite{lb4omptpds}. 
	\hl{In prior \mbox{work~\cite{lb4omptpds}}, we also instrumented other loops in SPHYNX. 
		However, we found that optimizing them yielded minimal performance gains due to their short runtime and already balanced workloads. 
		Thus, in this work, we focus only on the \textit{gravity} loop (L0).
		SPHYNX exemplifies large-scale real-world simulations with dominant computational hotspots where focused optimization can be highly effective.}

	\subsection{Performance Comparison of Oracle, Expert- and RL-based Scheduling Algorithm Selections}\label{sec:perfsel}

\begin{figure}[!htb]
		\centering
		\includegraphics[width=0.7\linewidth]{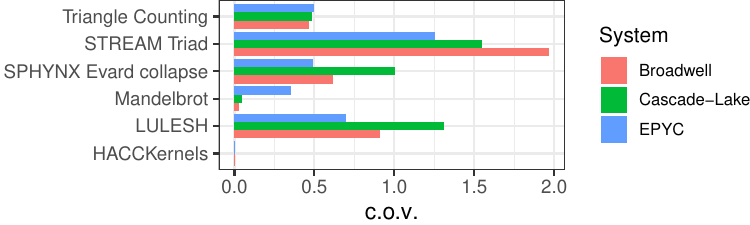}	    
		
		\caption{
				Coefficient of variation (\cov) per application-system pair considering the standard deviation of loop(s) execution time with every scheduling algorithm and chunk parameter divided by the average execution time of the loop(s). 
				High c.o.v. values indicate that the applications' loops performance is highly sensitive to the choice of scheduling algorithms.
		}\label{fig:covScheduling}
	\end{figure}

	We start by investigating how the choice of a scheduling algorithm affects the performance of applications' loops on every system. 
	For that, we calculate the coefficient of variation of loop execution times using every scheduling algorithm in LB4OMP for every application-system pair, as shown in Fig.~\ref{fig:covScheduling}.
	
	\begin{figure*}[!htb]
		\centering
        \makebox[\textwidth][c]{
		\includegraphics[width=1\linewidth]{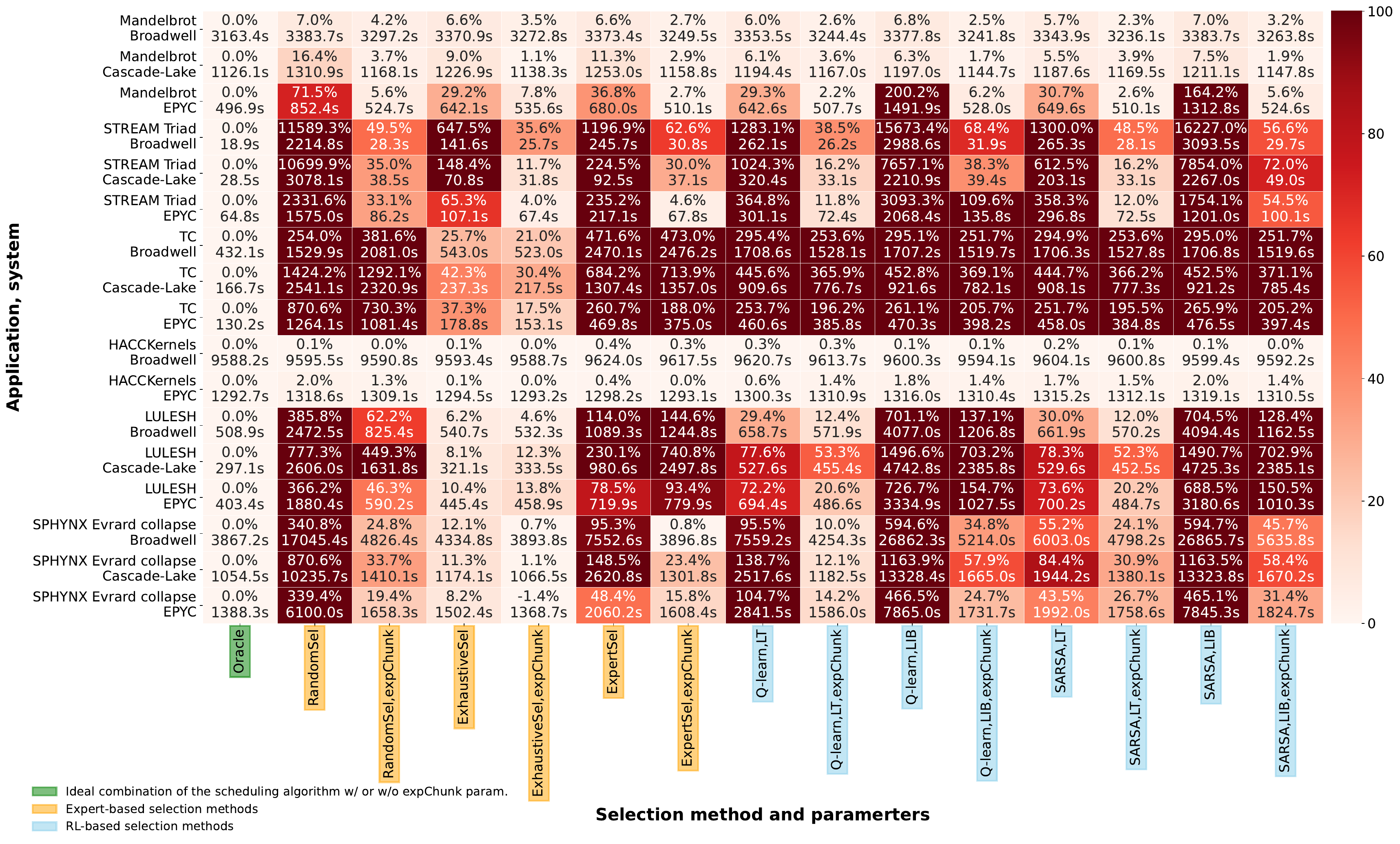}
		}
		\caption{
			Comparison of \highlightyellow{expert-} and \highlightblue{RL-based} scheduling algorithm selection methods in terms of performance degradation (\%) relative to \highlightgreen{\gt}.  
			The $x$-axis represents algorithm selection methods with or without the \expchunk parameter and reward type (LT or LIB), while the $y$-axis lists application-system pairs. 
			Shaded cells indicate relative performance to \gt (darker shades mean greater loss), and cell values represent execution times (in $s$) of applications, executing on systems, using the selected algorithms. 
			Performance degradation exceeds in certain cases \textcolor{white}{\highlightbrown{100\%}}, but the heatmap color gradient is restricted to 0\%-100\% for clarity.
		} \label{fig:allheatmap}
	\end{figure*}

	\streamtriad{} and \lulesh{} exhibit high \cov, sometimes exceeding 1 (Fig.\ref{fig:covScheduling}, \streamtriad{} on all systems and \lulesh{} on \gpu{} nodes). 
	This indicates a significant execution time disparity among scheduling algorithms, with a standard deviation larger than the mean. 
	This occurs because certain dynamic scheduling algorithms, like \ssdynamic, cause loss of data locality and increased synchronization overhead due to frequent scheduling rounds~\cite{lb4omptpds}.
	Consequently, algorithm selection methods are less effective for \streamtriad{} and \lulesh, where \static{} scheduling generally achieves the highest performance.

	\hacc{} shows \cov{} close to 0 (Fig.~\ref{fig:covScheduling}) because its compute-bound and balanced loop yields similar execution times across all scheduling algorithms. 
	Thus, \hacc{} may not need automated selection of the scheduling algorithm.

	\sphynx, \tc, and \mandel{} exhibit a \cov{} between 0.3 and 1 (Fig.~\ref{fig:covScheduling}), indicating a significant variance in execution times based on the scheduling algorithm used while presenting fewer outliers compared to \streamtriad{} and \lulesh{}. 
	\hl{Thus, these applications have the greatest need for automated scheduling algorithm selection and present a challenging scenario, as more algorithms may achieve high performance, increasing the selection complexity.}

	We investigated the performance of the automated scheduling algorithm selection methods compared to \gt (see Sect.~\ref{subsub:oracle} for the definition of \gt) in Fig.~\ref{fig:allheatmap}, which shows the performance overview of all methods. 
	The values in each cell report the median of 5 repetitions of the experiments. 
	\hl{\textbf{The complete dataset and additional results are available \mbox{online~\cite{rl4omp-data-zenodo}.}}}

	Using \expchunk{} generally improved performance across all methods. 
	For example, \sphynx{} on \gpu{} saw a reduction in performance degradation from 138.7\% to 12.1\% compared to \gt{} when \expchunk{} was used with \qlearn{} and LT reward. 
	Similarly, \streamtriad{} on \amd{} experienced a performance decrease of 358.3\% versus 12.3\% with \expchunk{} when using \sarsa{} with LT reward. 
	These results show that combining \textit{expert-knowledge} (e.g., via \expchunk) with RL-based methods can yield even higher performance than RL-based approaches exclusively.

	\qlearn{} and \sarsa{} with LT reward and \expchunk{} achieved similar performance. 
	For \lulesh{} on \gpu, \qlearn{} and \sarsa{} showed a maximum performance degradation of 53.5\% and 52.3\%, respectively. 
	\tc{} is an exception, where only \autoEXT{} with \expchunk{} came close to \gt. 
	This is due to the sparse graph of \tc{} that causes extreme load imbalance, with large chunks of iterations assigned to a few threads. 
	For \tc{}, high performance was only achieved by \ssdynamic{} and \static~\textbf{with} \textbf{\expchunk}, and only \autoEXT{} with \expchunk{} efficiently selected these.

	RL-based selection methods using LIB as a reward performed poorly. 
	Scheduling algorithms that minimize LIB often incur high overhead, with \ssdynamic{} and a chunk of 1 loop iteration being an extreme example. 
	LIB rewards favor such algorithms, resulting in low performance (see Sect.~\ref{sec:selection} for more details). 
	LIB only occasionally outperforms LT in compute-bound applications with high load imbalance, such as \mandel{} on \gpu, where \qlearn{} with \expchunk{} and LIB as a reward outperformed LT by 1.9\%. 
	As anticipated, prioritizing LIB as a reward is generally unsuitable for current systems, as the pursuit of minimal load imbalance leads to excessive scheduling overhead.

	\autoEXT{} with \expchunk{} achieved the performance closest to \gt{} in all cases, within 35\%. 
	In particular, for \sphynx{} on the \amd{} nodes, \autoEXT{} with \expchunk{} even surpassed \gt{} by 1. 4\%, probably due to small variations in execution time (approximately 2\%). 
	Since \gt{} is based on separate executions of different scheduling algorithms, automated selection methods may sometimes perform slightly faster than \gt. 
	This highlights the value of automated methods, as performance variability, even in homogeneous systems such as \amd{} nodes, can affect load balance and the selection of the optimal scheduling algorithm during execution.

	\subsection{Comparison of Oracle, Expert- and RL-based Scheduling Algorithm Choices}\label{sec:selection}

		

\begin{figure*}[!htb]
		\centering
        
		\includegraphics[trim=0cm 6.8cm 0cm 0cm,  clip=true, width=\linewidth]{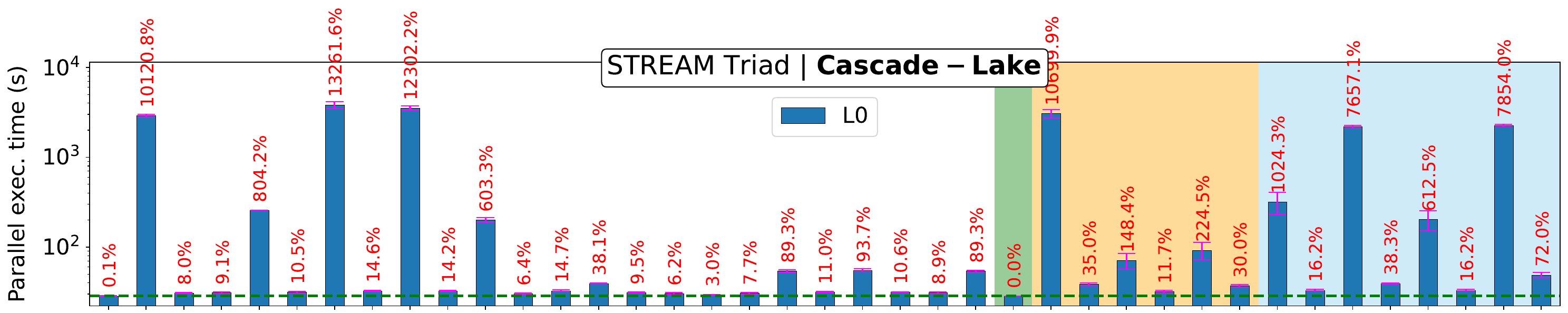}
		
		\includegraphics[trim=0cm 0cm 0cm 0cm,  clip=true, width=\linewidth]{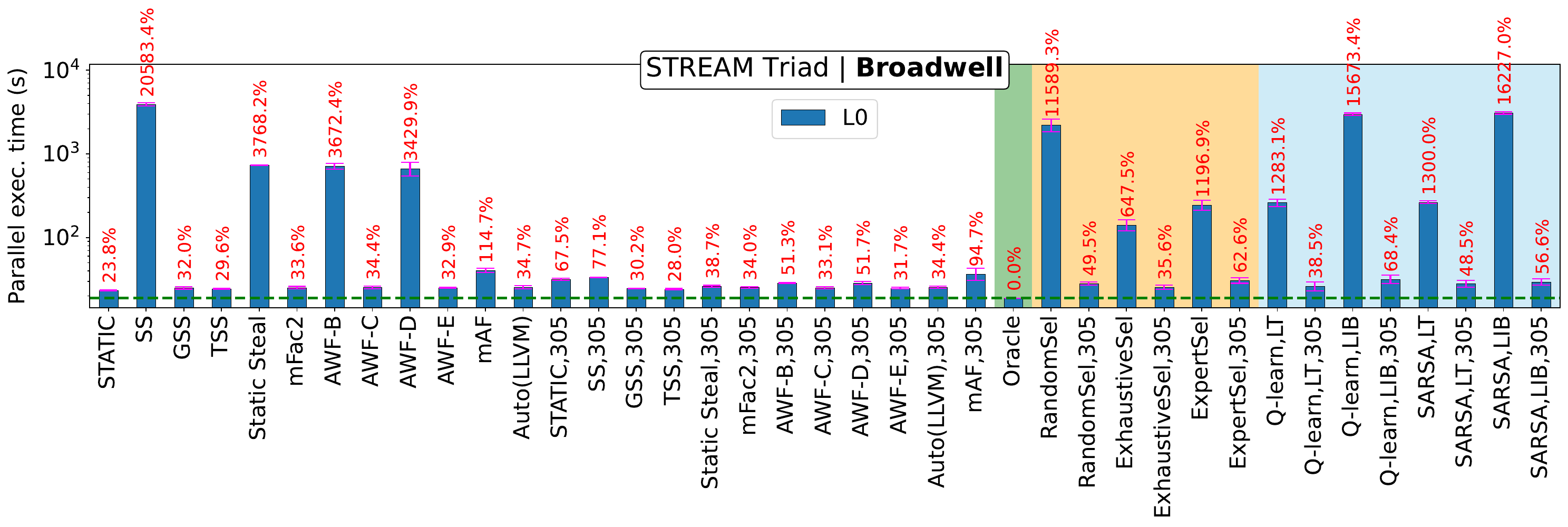}
		
		\includegraphics[trim=0cm 8cm 0cm 0.2cm,  clip=true, width=\linewidth]{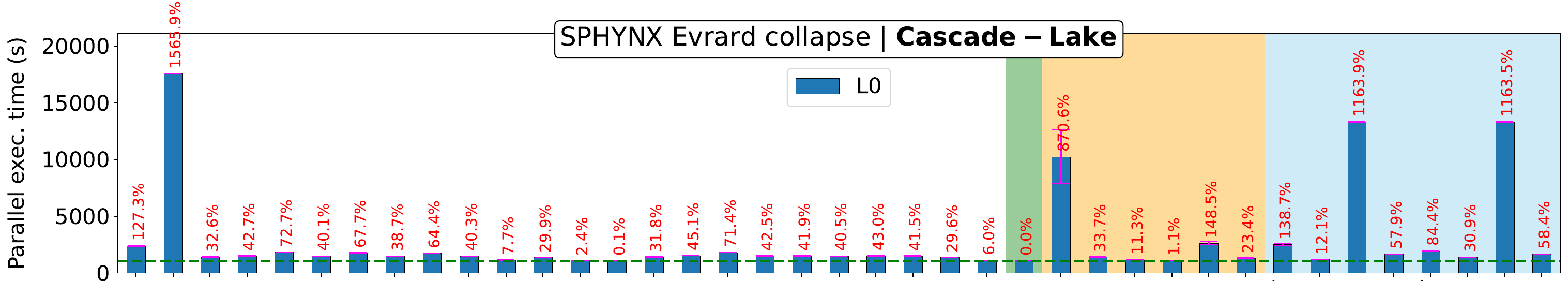}
		
		\includegraphics[trim=0cm 0cm 0cm 0.5cm,  clip=true, width=\linewidth]{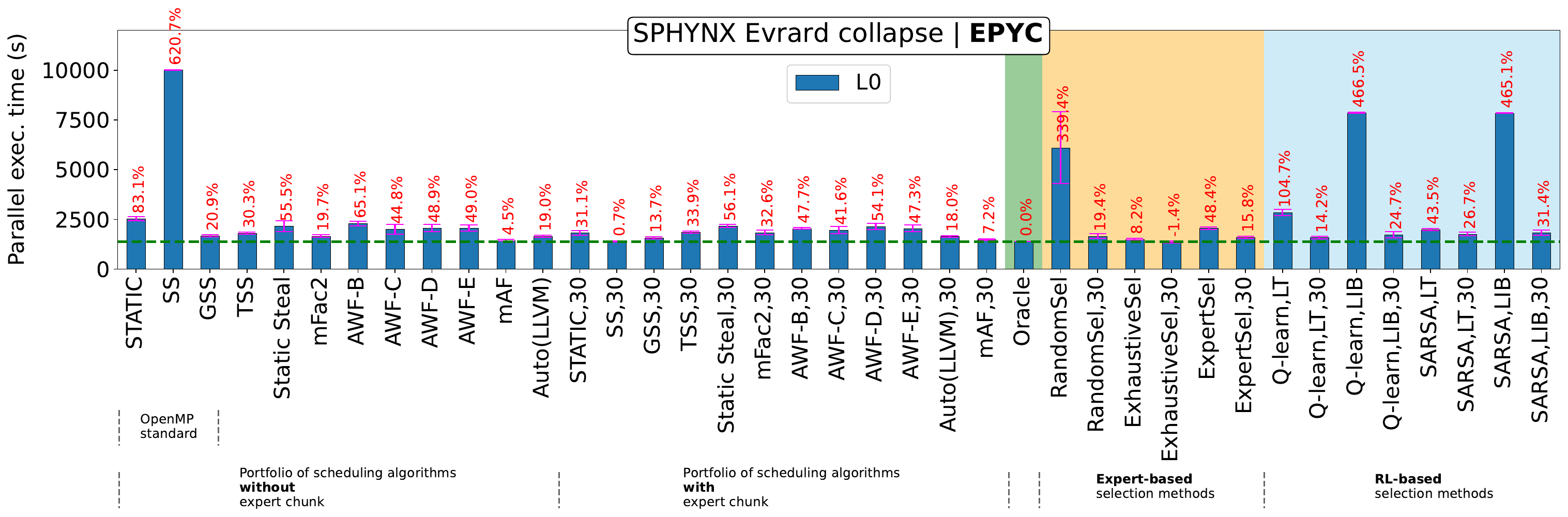}
		\vspace{0.3cm}\caption{
			Median parallel loop(s) execution time (s), top two plots show results for \streamtriad{} on \gpu{} and \xeon{} while the bottom two plots show results for \sphynx{} executing on \gpu{} and \amd.
			The $x$ axis shows the scheduling algorithms (left up to center) and selection methods with parameters (right).
			The $y$ axis shows the parallel execution time (in $s$).
			Percentages (\%) denote performance degradation relative to \highlightgreen{\gt}.
			The highlighted regions denote: in white (left) the use of individual scheduling algorithms, followed by \highlightgreen{\gt}, the \highlightyellow{Expert-based selected}, and the \highlightblue{RL-based selected} scheduling methods. 
			The \textcolor{darkgreen}{horizontal dotted line} represents the baseline \highlightgreen{\gt}~across all other scheduling algorithms and selection methods. 
			The \textcolor{magentaFig3}{error bars} denote the standard deviation of the 5 experiment repetitions.
		}
		\label{fig:perfSphynx}
	\end{figure*}

	To understand the selection and performance achieved by the different automated selection methods, we first focus on more detailed results that show the performance achieved by all scheduling algorithms and selection methods for \streamtriad, the top two graphs of Fig.~\ref{fig:perfSphynx}, and \sphynx, the bottom two graphs of Fig.~\ref{fig:perfSphynx}. 
	We selected \streamtriad{} to deliberately highlight the challenges behind scheduling algorithm selection on scenarios where a single wrong choice can cause extreme performance degradation. 
	\sphynx{} was selected to highlight the potential of scheduling algorithm selection on actual real-world applications.

	In Fig.~\ref{fig:perfSphynx}, the results for \sphynx{} show that certain scheduling algorithms cause significant overhead and loss of data locality. 
	For example, \sphynx{} executed with \ssdynamic{} and \static{} without \expchunk{} had execution times higher than 600\% and 80\% than when using \gt, respectively. 
	\ssdynamic{} without \expchunk{} leads to extreme overhead and loss of data locality, while \static{} results in high execution times due to the imbalanced workload of the \textit{gravity} (L0) loop of \sphynx.
	
	It is also important to highlight that the \textit{gravity} (L0) loop of \sphynx{} evolves over time steps showing different workload and workload imbalance over the course of the execution.
	This poses another challenge for automated selection methods, as different time-steps can influence which algorithms are selected, potentially leading to indecision. 
	
	The results shown in Fig.~\ref{fig:perfSphynx} also clarify why all automated selection methods benefit from the \expchunk{} parameter as it reduces the overhead and locality issues of most dynamic scheduling algorithms. 
	Consequently, this fact also reduces the overhead of the exploration phase of automated methods. 
	
	\begin{figure*}[!htb]
		\centering
		\includegraphics[width=\linewidth]{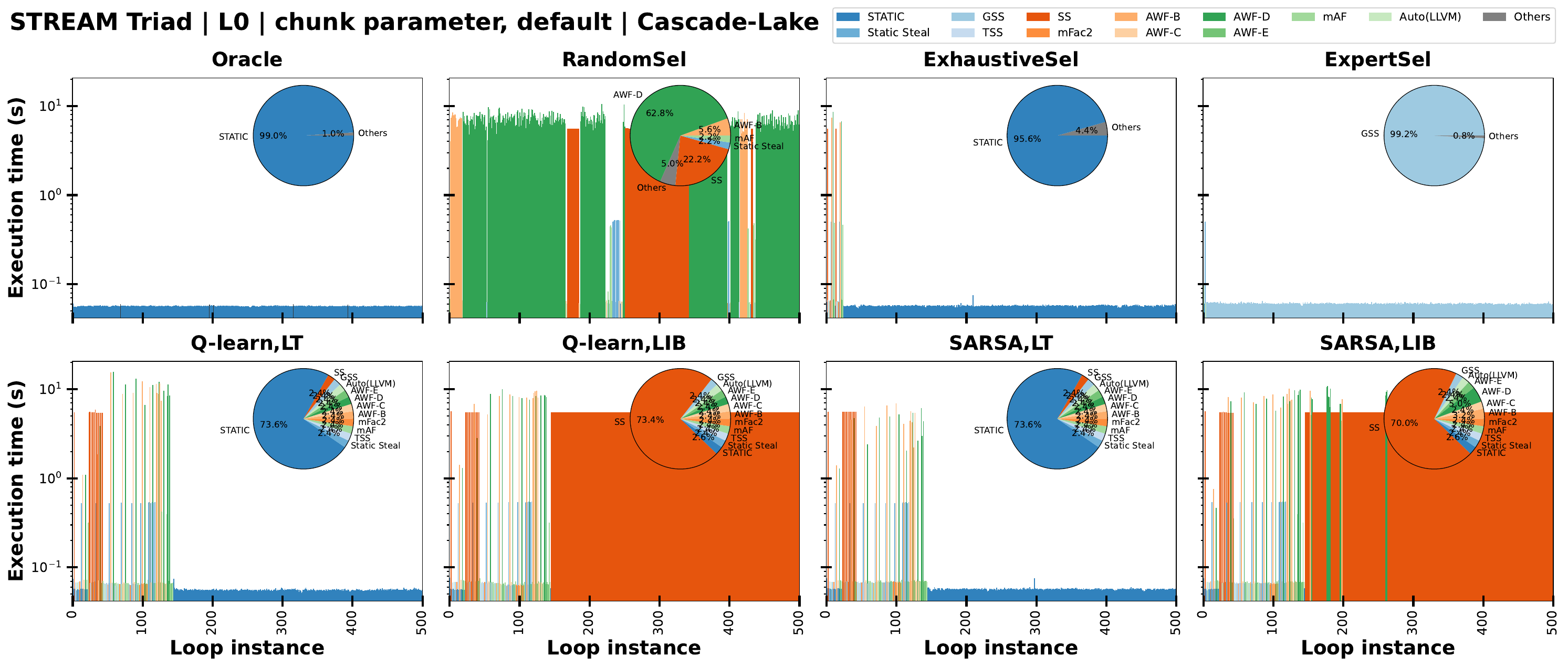}	
		\rule{\textwidth}{0.4pt}
		\includegraphics[width=\linewidth]{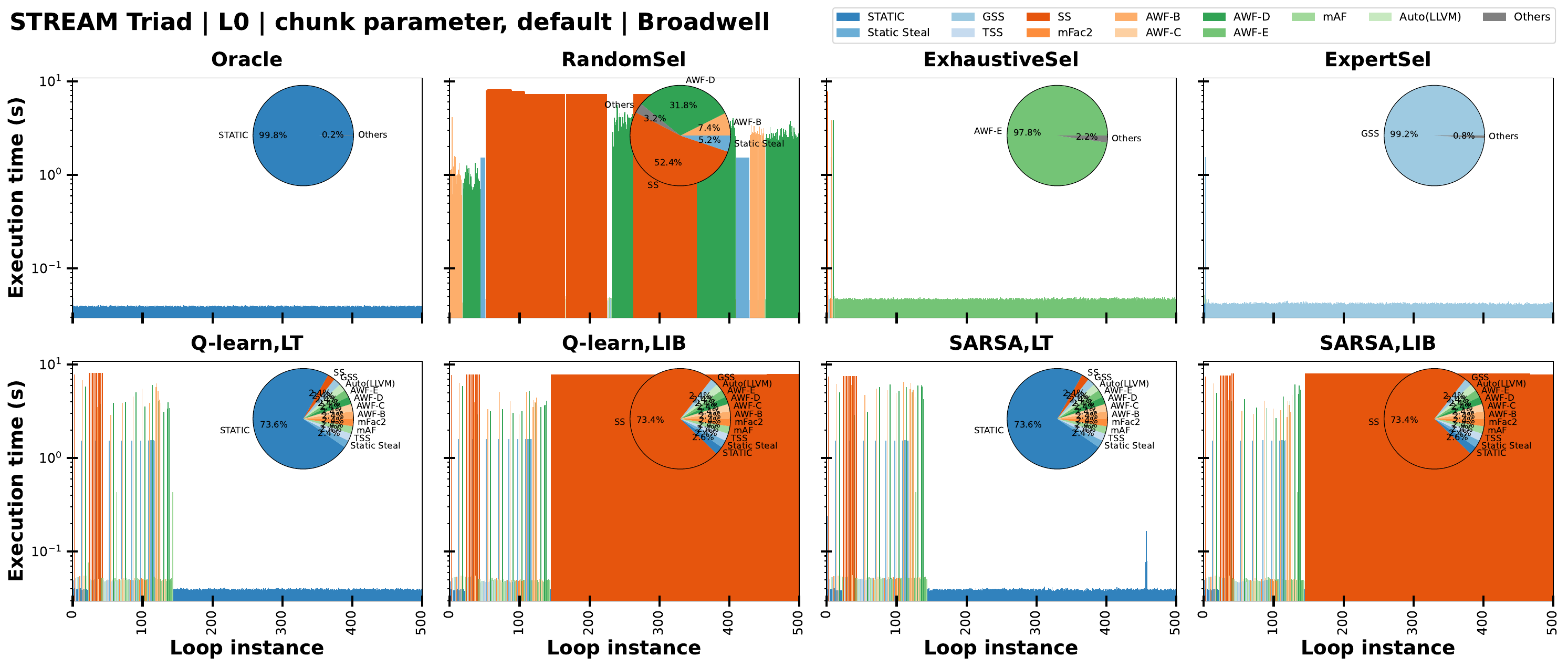}
		\caption{
			Scheduling algorithms selected by the expert- and RL-based automated selection methods or by \gt{} per loop instance for \streamtriad{} executing on \gpu{} (top) and \xeon{} (bottom), both \textbf{without} \expchunk{} parameter.
			The height of each bar represents the execution time (s) of the selected scheduling algorithm for the given loop instance shown on the $x$-axis. 
			The pie charts show the percentage of times a given scheduling algorithm was selected. 
			The scheduling algorithms that were selected less than $2\%$ of the loop instances were aggregated under \textit{Others}.
			\hl{For the RL-based methods, $28.8\%$ of the time-steps (144 out of 500) is always spent on the learning process.}
		}
		\label{fig:selectionStream}
	\end{figure*}
	\begin{figure*}[!htb]
		\centering
		\includegraphics[width=\linewidth]{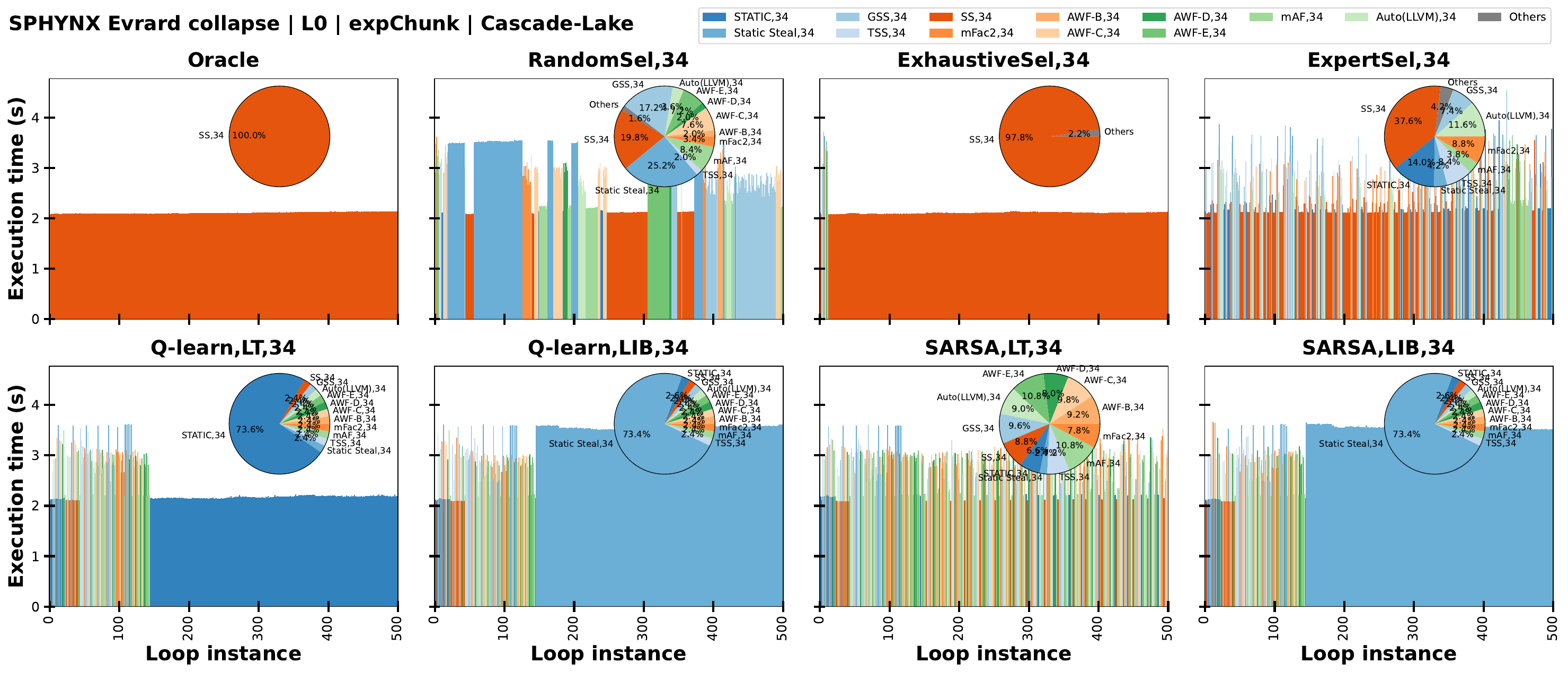}
		\rule{\textwidth}{0.4pt}
		\includegraphics[width=\linewidth]{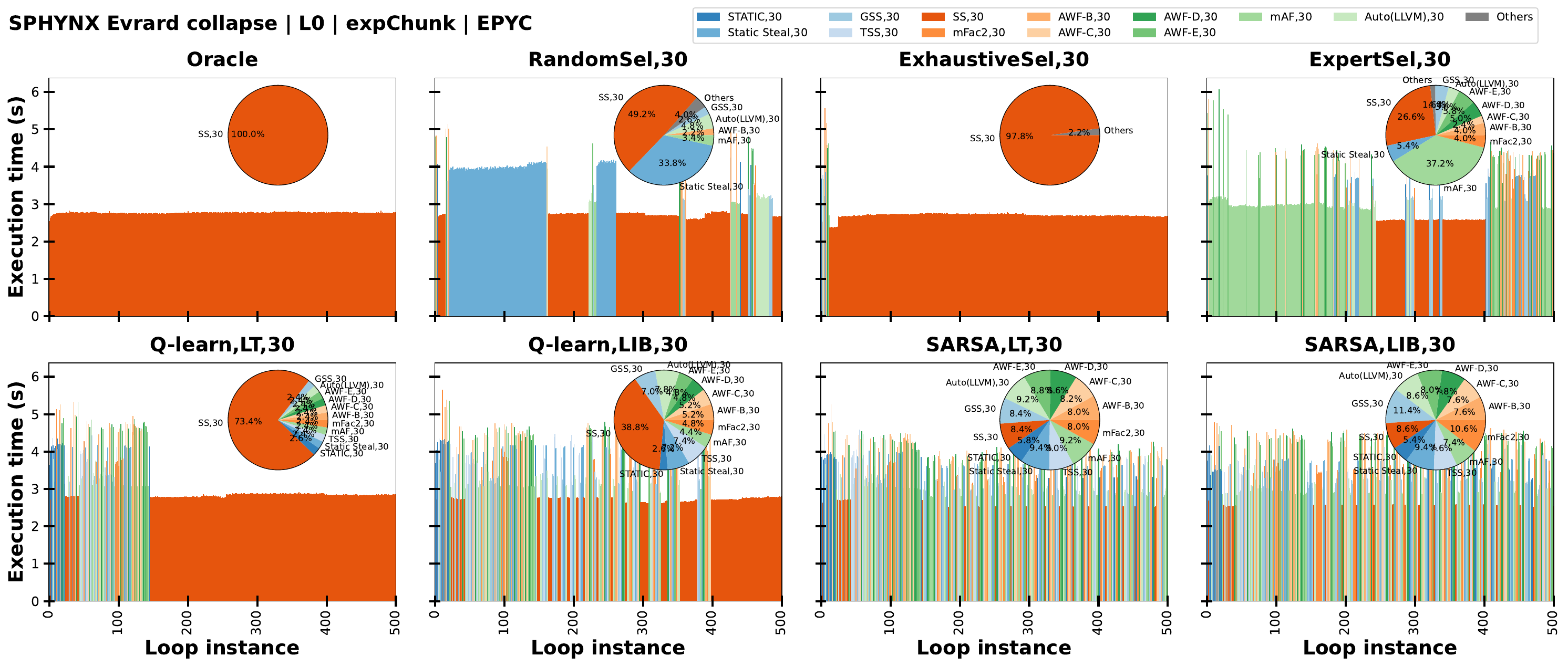}
		\caption{
			Scheduling algorithms selected by the expert- and RL-based automated selection methods or by \gt{} per loop instance for \sphynx{} executing on \gpu{} (top) and \amd{} (bottom), both \textbf{with} \expchunk{} parameter.
			The height of each bar represents the execution time (s) of the selected scheduling algorithm for the given loop instance shown on the $x$-axis. 
			The pie charts show the percentage of times a given scheduling algorithm was selected. 
			The scheduling algorithms that were selected less than $2\%$ of the loop instances were aggregated under \textit{Others}.
			\hl{For the RL-based methods, $28.8\%$ of the time-steps (144 out of 500) is always spent on the learning process.}
		}
		\label{fig:selectionSphynx}
	\end{figure*}

	For \streamtriad, \ssdynamic, \steal, \awfb, and \awfd{} without \expchunk{} achieved execution times of orders of magnitude higher than \gt{} (\static{} in this case). 
	\streamtriad is an purely memory-bound code and any dynamic scheduling technique that requires numerous scheduling rounds will cause extreme overhead and data locality loss. 
	
	Another aspect of \streamtriad{} (Fig.~\ref{fig:perfSphynx}) is that for both computing systems, \static{} \textbf{without} an \expchunk{} parameter outperforms \static{} \textbf{with} \expchunk. 
	This creates an even harder challenge for automated scheduling algorithm selection. 
	On the one hand, in the case of \streamtriad, the highest performance is achieved \textbf{without} an \expchunk{} parameter. 
	However, not using the \expchunk{} parameter with automated selection methods makes their exploration phase more costly. 
	Thus, one must decide between a costly exploration/learning phase that will lead to the optimal scheduling algorithm selection and a less costly exploration/learning phase that will not be able to reach the optimal selection. 
	
	\hl{
		When comparing the selection methods with standard OpenMP scheduling algorithms such as \static, \ssdynamic, and \gss, noticeable performance differences can be observed.
		Expert-based methods with the \expchunk{} parameter showed execution times lower than standard algorithms in multiple cases; for example, on the \gpu{} nodes, GSS is approximately $30\%$ slower than \autoEXT{} for the SPHYNX Evrard collapse (\mbox{Fig.~\ref{fig:perfSphynx}}).
		RL-based methods also outperform GSS in some cases; specifically, \qlearn{} with LT reward and \expchunk{} outperforms GSS by $20\%$ for the same SPHYNX scenario on \gpu{} nodes (\mbox{Fig.~\ref{fig:perfSphynx}}).
		In general, the OpenMP standard scheduling algorithms' options are typically at least about $10\%$ slower than \gt, with performance degradation reaching a minimum of $30\%$ in numerous cases. \textbf{These highlight the importance of extending OpenMP with additional scheduling options and selection methods capable of identifying the highest-performing algorithm, or combination thereof, for a given execution context.}
	}
	
	To understand and expose the selected scheduling algorithms by every selection method, we examine each method's selections, highlighting the costs of their exploration/learning phases and their ability to find the highest-performing scheduling algorithm. 
	Fig.~\ref{fig:selectionStream} shows the selections of scheduling algorithms by the automated selection methods based on expert, RL and by \gt, while executing \streamtriad{} on \gpu (top) and on \xeon (bottom), while Fig.~\ref{fig:selectionSphynx} shows the scheduling algorithms selected by expert- and RL-based selection methods using \expchunk, and by \gt, while executing \sphynx{} on \amd.

	Both Fig.~\ref{fig:selectionStream} and Fig.~\ref{fig:selectionSphynx} expose why RL-based selection methods while using the LIB reward achieve low performance. 
	For example, on \amd{} nodes that executed \sphynx{} (Fig.~\ref{fig:selectionSphynx}), \sarsa{} and \qlearn{} using the LIB reward were unable to find a selection even after the algorithm learning phase. 
	In other cases, for example \streamtriad{} executing on \gpu{} or \xeon{} (Fig.~\ref{fig:selectionStream}), the RL-based methods found the scheduling technique that achieved the lowest LIB (\ssdynamic), yet at the cost of high scheduling overhead.
	
	For \streamtriad, Fig.~\ref{fig:selectionStream}, the RL-based methods using the LT reward achieved the same selection as \gt{} after their learning phase in both systems; see Fig.~\ref{fig:selectionStream}. 
	\hl{\textit{This indicates that RL-based selection adapts very well to different environments by selecting the highest-performing scheduling algorithm after the learning phase.}}
	Nevertheless, we observe that a single instance with a costly scheduling technique such as \ssdynamic{} can lead to loop instance execution orders of magnitude longer than others. 
	This shows that RL-based methods must incorporate expert knowledge, such as using \expchunk, or the executing application must encompass numerous loop instances (time steps) to counterbalance the performance degradation associated with the learning phase. 
	\hl{For example, the 144 loop instances required for the learning phase account for $28.8\%$ of the total 500 loop instances of \streamtriad.
		The cost of this learning phase outweighs the benefits gained by taking advantage of the selection of an optimal scheduling algorithm afterward. 
		Thus, this cost could be compensated for by a sufficiently large number of loop instances.}
	
	\hl{The exploration phase of expert-based methods is much shorter (less than $10\%$ of the total loop instances), as shown in \mbox{Fig.~\ref{fig:selectionStream}}~and \mbox{Fig.~\ref{fig:selectionSphynx}}.}
	However, for \streamtriad{} in Fig.~\ref{fig:selectionStream}, we can observe that although the scheduling algorithms selected by \autoEXT{} and \autoEXP{} achieved high performance, they did not select the same algorithms as \gt{} for both systems (\gpu{} and \xeon). 
	This means that these selection methods reached a ``suboptimal'' selection of scheduling algorithms for \streamtriad, leaving opportunities for performance improvement.
	
	For \streamtriad{} executed on \gpu, \autoEXT{} reached the same ``optimal selection'' as \gt. 
	However, \autoEXT{} arrived at a different selection than \gt{} when executing \streamtriad{} on \xeon. 
	\textit{This is an indication that expert-based selection methods are not always capable of finding the highest-performing algorithm when executing on different environments while RL-based methods adapt to such changes.}
	
	The results of \sphynx{} executed on \gpu{} using \qlearn{} with the LT reward (see the top of Fig.~\ref{fig:selectionSphynx}) highlight the importance of automated selection methods. 
	In this scenario, \qlearn{} identified \static{} with \expchunk{} as the highest-performing scheduling algorithm, while \gt{} suggested that \ssdynamic with \expchunk{} would be the best choice.
	To understand this discrepancy, we refer to the performance of \sphynx{} on \gpu{} with \static{} and \ssdynamic{} using \expchunk, as shown in Fig.~\ref{fig:perfSphynx}. The performance difference between the two techniques is minimal, just 2. 4\% in favor of \ssdynamic. 
	However, such a small margin, combined with potential system variations, led to a scenario in which \static{} with \expchunk{} outperformed \ssdynamic{} with \expchunk{} during the \qlearn{} experiments with the LT reward, resulting in \static{} being selected.

	From the results for \sphynx{} that execute on the \amd{} nodes (bottom of Fig.~\ref{fig:selectionSphynx}), we see that \qlearn{} with the LT reward matched the selection of \gt{} after the learning phase. 
	However, \sphynx{} executed with \sarsa, LT reward, and \autoEXP{} failed to find the best algorithm (in both systems). 
	This indicates that these methods tend to `overreact' to evolving loop behavior -- the modified loop from \sphynx{} calculates \textit{gravity} over particles and the precise operations evolve across time-steps depending on the particle distribution. 
	For \sarsa, fine-tuning parameters such as the learning rate ($\alpha$) and its decay may be required. 
	Another possibility of fluctuating selection by \sarsa{} is that the use of \expchunk{} makes most scheduling algorithms achieve comparable performance, which can influence the learning process. 
	More research is needed to understand the effects of combining expert knowledge (e.g, via an \expchunk) and RL for the selection of scheduling algorithms. 
	
	Lastly, Fig.~\ref{fig:selectionStream} and Fig.~\ref{fig:selectionSphynx} clarify why \autoEXT{} outperformed in most cases. 
	It matched \gt's selection, after trying each algorithm once, which, for \streamtriad{} the selection overhead was low and the selected scheduling algorithms (STATIC or AWF-E), although not optimal, performed well, and for \sphynx, incurred minimal selection overhead and identified the highest-performing algorithm (SS,34 and SS,30).

	\section{Discussion}\label{sec:discussion}
	
	
	Addressing \textbf{RQ1} of how expert- and RL-based selection methods compare, \textbf{the primary benefit of RL-based methods lies in their adaptability to various environments, although they require a longer learning phase, resulting in higher selection overhead compared to expert-based methods.} 
	This flexibility makes RL-based methods more suitable for evolving system architectures. 
	
	In contrast, expert-based methods leverage expert/domain knowledge to reduce algorithm selection costs and enhance performance. 
	Expert-based selection methods excel 
	in scenarios where the exploration cost is prohibitive or where sufficient domain knowledge is available to effectively guide the selection process. 
	These methods provide consistent, low-overhead performance, making them particularly suitable for applications that involve memory-bound loops or little workload imbalance. 
	However, their reliance on fixed rules and parameters necessitates modifications and re-experimentation for future systems, limiting their adaptability to novel or rapidly evolving workload patterns and system architectures.
	
	RL-based methods show significant potential to adapt to diverse and dynamic environments. 
	These methods leverage runtime data to refine decision-making processes, enabling improved performance over time, particularly in compute-bound scenarios, highly variable applications, and emerging ``unknown'' systems. 
	While the exploration phase incurs a notable performance cost, this impact diminishes as the system accumulates knowledge. 
	The adaptability of RL-based methods makes them a robust choice for heterogeneous systems. 
	
	\hl{Future research could focus on mitigating exploration costs by integrating pre-trained models or hybrid approaches that combine expert knowledge with reinforcement learning.} 
	\hl{\textit{The data from this study can be used to train models for model-based RL techniques}, such as \mbox{I2A~\cite{i2a2017}} and Prioritized \mbox{Sweeping~\cite{corneil2018efficient}}, potentially reducing or eliminating learning overhead while retaining the adaptability of RL-based methods.}
	Additionally, this work lays the foundation for integrating scheduling algorithm selection at multiple levels, such as when combining OpenMP scheduling with MPI-level scheduling strategies to optimize performance across the entire parallelism hierarchy.
	
	
	Regarding which parameters influence RL-based selection considered by \textbf{RQ2}, several parameters significantly influence the selection and performance of RL-based selected scheduling algorithms. 
	Key findings include:
	(1)~\textbf{The choice of reward strategy is crucial}. Using \texttt{LIB} as a reward often leads to poor performance by favoring algorithms with high overhead costs.
	In particular, the \texttt{LIB} reward was selected to emphasize that focusing solely on mitigating load imbalance does not necessarily result in performance gains. 
	Counterintuitively, it has recently been shown that introducing a controlled degree of imbalance may be beneficial to improving memory usage~\cite{Afzal2023}.
	(2)~\textbf{The \texttt{explore-first} strategy is particularly costly}. Future research aims to investigate alternative exploration techniques to address this limitation.
	(3)~\textbf{\sarsa{} may take longer or even fail to identify an appropriate scheduling algorithm, while \qlearn{} typically makes a selection immediately after the learning phase}.
	This analysis reveals the importance of reward and exploration strategies, as well as the specific RL algorithm used, in optimizing scheduling performance.
	
	
	\hl{In this work, \textbf{we integrate ``expert knowledge'' through the \expchunk{} parameter with RL-based algorithm selection methods and show that this approach achieves superior performance compared to RL-based selection methods without \expchunk}, e.g. 25-40\% for \mbox{memory-bound} loops such as \streamtriad{} (see \mbox{Fig.~\ref{fig:allheatmap})}.}
	This integration of expert knowledge with RL strategies reduces online selection overhead, enhances selection accuracy, and improves application performance, partially addressing \textbf{RQ3}. 
	
	Another key factor influencing the performance of RL-based methods is the size of the scheduling algorithm portfolio. 
	A larger portfolio prolongs the learning phase, thus increasing associated costs. 
	While the size of the \textit{expert} portfolio considered in this work (12 algorithms)~\cite{auto4omp}, remains manageable for expert-based methods, it substantially increases the learning phase costs for RL-based selections.

	The selection methods of expert- and RL-based scheduling algorithms rely on time-stepping applications (multiple executions of the same loop(s)) to learn effectively through the exploration and exploitation phases. 
	While this may seem limiting, these methods can also benefit single-sweep applications by iteratively executing them on a specific system and recording their performance for each scheduling algorithm. 
	This iterative process enables the identification of the optimal scheduling algorithm for future executions of the same application on the same system, offering valuable insights to maximize application performance.
	
	\hl{
		In summary, the experimental analyses indicate that no single scheduling algorithm or selection strategy consistently delivers the highest performance across all tested application-system combinations. 
		Expert-based selection methods tend to perform reliably with low overhead.
		Reinforcement learning (RL)-based methods, on the other hand, exhibit greater adaptability at a much higher overhead. 
		The expert chunk parameter further enhances both expert- and RL-based approaches by reducing scheduling overhead and improving data locality. 
		Finally, standard OpenMP scheduling algorithms, such as \gss, \static, and \ssdynamic, often lead to suboptimal performance, often $30\%$ or more below that of the \gt{} baseline, especially for applications with irregular loop behavior or evolving imbalance patterns such as \sphynx.
	}
	
	\hl{
		The current \textit{high overhead of RL-based methods limits their practical applicability} in short loops or applications with limited iteration counts. 
		\textit{To improve the usability of RL-based selection methods, future work should focus on reducing exploration costs, for example, by incorporating prior knowledge, enabling transfer \mbox{learning~\cite{transferLearning2010}}, or developing model-based RL techniques\mbox{~\cite{modelbasedSurvey2023}}.} 
		A \mbox{model-based} approach could leverage previously collected performance data to guide selection decisions more efficiently, potentially eliminating the need for costly exploration phases. 
		These improvements would significantly increase the viability of RL-based scheduling in real-world HPC applications.
	}
	
	\hl{
		This selection reflects a significant diversity in architecture, vendor characteristics, and core configurations, helping to demonstrate the consistency of the results across systems and hardware generations.
		However, validating the results on newer hardware (e.g. Intel Sapphire Rapids or AMD Genoa) would further strengthen the conclusions, and this remains a valuable direction for future work.
	}

	\hl{
		Finally, it is important to recognize that there is no universal solution to the scheduling algorithm selection problem. 
		The \mbox{"no-free lunch"} \mbox{theorem~\cite{nofreelunch1997, nofreelunch2020}} highlights that the effectiveness of any algorithm depends on the specific characteristics of the task and environment. 
		In the context of scheduling, both the task (workload) and environment (underlying system, operating system, runtime, etc.) are dynamic, making it impossible to identify a single, universally optimal solution. 
	}
	
	\section{Related Work}\label{sec:rlw}

	The selection of scheduling algorithms is an instance of the algorithm selection problem proposed by Rice~\cite{RICE1976}. 
	Various selection methods for scheduling algorithms have been proposed in the past, including decision trees~\cite{RuntimeEmpSched}, exhaustive offline search~\cite{Sreenivasan2019frameworkautotuning}, simulation-assisted selection~\cite{DLS4LB}, machine learning-based selection~\cite{banicescu:2013:a, Boulmier:2017a}, and expert-based selection~\cite{auto4omp}.
	
	For executing OpenMP multithreading applications on symmetric multiprocessor~(SMP) architectures, threads could be organized across two levels based on physical cores and hyperthreads within a physical core~\cite{RuntimeEmpSched}. 
	Three empirical methods were proposed to select the optimal number of threads and scheduling algorithms from a set of five options: \static~\cite{staticSchedules}, \ssdynamic~\cite{SS}, \gss~\cite{GSS:1987}, and \tss~\cite{TSS:1993}, affinity~\cite{markatos1994using}. 
	These methods require precomputed information and, once selected, a scheduling algorithm persists throughout the entire execution of the application. 
	
	Another approach~\cite{thoman2012automatic} used the polyhedral model~\cite{benabderrahmane2010polyhedral}, compiler analysis, and a load-aware run-time system. 
	The focus was on selecting the algorithm that performed the best among the OpenMP standard-compliant options (i.e. \static, \ssdynamic{} and \gss). 
	This approach required specific compiler analysis information. 
	
	Another framework determines the best-performing scheduling algorithm for OpenMP loops~\cite{Sreenivasan2019frameworkautotuning}.
	The framework considers \static, \ssdynamic as scheduling methods, and \{1, 8, 16\} as chunk parameter values.
	The framework automatically generates several instances of the same loop with variable parameters \texttt{kind} and \texttt{chunk} of the \texttt{schedule(kind, chunk)} clause and thread configurations. 
	The selection process is then performed offline through extensive experimentation. 
	
	\hl{
		Algorithm selection has also recently proven effective in \mbox{domain-specific} contexts. 
		\mbox{Newcome et al.~\cite{newcome2025mdselection}} apply it to \mbox{short-range} molecular dynamics simulations, showing that runtime kernel choices can significantly improve performance across configurations. 
		While their work focuses on a specific application domain, in this work we address a broader problem: scheduling algorithm selection in OpenMP across different applications and systems.
	}
	
	\hl{
		Another recent study by Silva \mbox{et al.~\cite{silva2025autotuning}} presents an example where the authors apply auto-tuning techniques to optimize OpenMP dynamic scheduling for a \mbox{real-world} seismic imaging application based on Full Waveform Inversion (FWI). 
		Their work demonstrates the practical impact of tuning chunk sizes in a single application to improve runtime performance across problem sizes and architectures. 
		In contrast, the present work generalizes beyond chunk tuning for a single algorithm by investigating the automated selection of scheduling algorithms themselves, spanning a portfolio of 12 techniques, across diverse applications and systems. 
		Moreover, in this work we compare and combine expert-based and RL-based selection strategies, evaluating runtime adaptability and performance of the different strategies.
	}
	
	Compared to the related work discussed above, the RL-based algorithm selection methods introduced in this study select scheduling algorithms from a significantly larger and more advanced portfolio of options, leveraging the LB4OMP library. 
	A key advantage of these RL-based algorithm selection methods is that they do not require prior information about the application or loop, unlike recent approaches~\cite{RuntimeEmpSched, thoman2012automatic, Sreenivasan2019frameworkautotuning}. 
	In addition, RL-based algorithm selection methods continuously learn during execution, allowing them to adapt to unpredictable system noise and/or dynamic workload changes.
	Furthermore, this work stands out by conducting a comprehensive comparative study of the performance, strengths, and weaknesses of expert- and RL-based methods and exploring the potential benefits of combining these algorithm selection methods.

	\section{Conclusion and Future Work}\label{sec:conclusion}
	
	This work studied the problem of learning to select scheduling algorithms in OpenMP. 
	We considered expert-based (\autoEXT{} and \autoEXP) and RL-based (\qlearn and \sarsa) algorithm selection approaches and compared them against \gt{} and \autoRND{} selections. 
	We evaluated the effectiveness of each approach through an extensive performance analysis campaign with six applications and three systems. 
	We showed that expert- and RL-based approaches are comparable in terms of selections and performance, namely that RL-based selections deliver the highest-performing selections but at higher exploration cost (due to model-free \qlearn and \sarsa), while expert-based selections deliver less performing selection but at reduced exploration cost. 
	
	RL-based scheduling algorithm selection could be improved by using model-based approaches (requiring training a selection model) that will greatly reduce the overload of learning during execution. 
	The results collected in this work can serve to train such a model.
	
	Both expert- and RL-based algorithm selection methods require time-stepping applications to enable learning through exploration and exploitation phases.
	However, these selection methods can also be effectively employed in single-sweep applications, as long as they are executed repeatedly on a particular system and the performance of each scheduling algorithm is recorded. 
	
	This study improves node-level scheduling within the broader multilevel scheduling problem\footnote{Multilevel scheduling project: \url{https://hpc.dmi.unibas.ch/research/mls/}}, which aims to leverage increased hardware parallelism and heterogeneity across various levels.

	Future research will extend these methods to automatically select scheduling algorithms for node-level heterogeneous architectures, such as CPUs and GPUs, and other levels of hardware parallelism, such as cross-node (with MPI). 
	This will include adaptations to handle unpredictable variations and potential failures in applications and resources during execution.
	In addition, conducting an ANOVA analysis on the collected data to identify the most relevant and influential factors in the selection of the automated scheduling algorithm will also help refine and optimize the decision-making process of the RL algorithms.

\bibliographystyle{acm}
\bibliography{main}
\end{document}